\documentclass[12pt]{article}
\pdfoutput=1
\usepackage{tikz}
\usetikzlibrary{matrix,arrows,positioning,shapes,snakes,fadings,decorations.pathmorphing,
decorations.pathreplacing,automata,shadows,decorations.markings,patterns}
\usepackage[thicklines]{cancel}
\usetikzlibrary{calc,shapes.callouts,shapes.arrows}
 
\usepackage{mathtools}
\usepackage{jheppub}
\usepackage{amsmath,amssymb,euscript,array,mathrsfs,appendix,ctable,marvosym,bbm}
\usepackage{arydshln}
\usepackage{todonotes}
\usepackage{graphicx}
\usepackage{float}
\usepackage[margin=1cm,font=small]{caption}
\usepackage[margin=.2cm,font=footnotesize]{subcaption}
\usepackage[normalem]{ulem}
\usepackage{cleveref}
\usepackage[bottom]{footmisc}

\def\IR{\mathbb{R}}

\newcommand\blank[1]{#1}
\renewcommand\blank[1]{}
\def\Buildrel#1\over#2\under#3{\mathrel{\mathop{\kern0pt
#2}\limits^{#1}_{#3}}}

\newcommand{\pp}{{=\!\!\!|}}

\usepackage{enumitem}

\def\Lax{{\mathscr L}}

\newcommand{\Tr}{\operatorname{Tr}}

\def\B0{{\boldsymbol 0}}

\def\RR{{\mathfrak R}}

\def\N{{\cal N}}
\def\J{{\cal J}}

\def\det{{\rm det}}

\def\ee{\boldsymbol{e}}
\def\RR{{\mathfrak R}}

\def\R{{\mathbb R}}

\def\Dbarslash{\,\,{\raise.15ex\hbox{/}\mkern-12mu {\bar D}}}
\def\Dslash{\,\,{\raise.15ex\hbox{/}\mkern-12mu D}}
\def\delslash{\,\,{\raise.15ex\hbox{/}\mkern-9mu \partial}}
\def\delbarslash{\,\,{\raise.15ex\hbox{/}\mkern-9mu {\bar\partial}}}

\def\be{\begin{equation}}
\def\ee{\end{equation}}

\renewcommand{\R}{\mathcal{R}} 
\renewcommand{\RR}{(\mathcal{R}^{2})} 

\title{Classical and Quantum Aspects of Yang-Baxter Wess-Zumino Models}
 \author[a]{Saskia Demulder,}
 \author[a]{Sibylle Driezen,}
\author[a,b]{Alexander Sevrin,}
\author[a,c]{\\ and Daniel C. Thompson}
\affiliation[a]{
Theoretische Natuurkunde, Vrije Universiteit Brussel\\  \& The International Solvay Institutes\\
Pleinlaan 2, B-1050 Brussels, Belgium }
\affiliation[b]{also at the Physics Department, Universiteit Antwerpen\\
Campus Groenenborger, 2020 Antwerpen, Belgium}
\affiliation[c]{Department of Physics, Swansea University\\
Singleton Park, Swansea SA2 8PP, U.K.}

\emailAdd{Saskia.Demulder@vub.be}
\emailAdd{Sibylle.Driezen@vub.be}
\emailAdd{Alexandre.Sevrin@vub.be}
 \emailAdd{D.C.Thompson@Swansea.ac.uk}

\abstract{We investigate the integrable Yang-Baxter deformation of the 2d Principal Chiral Model with a Wess-Zumino term. For arbitrary groups, the one-loop $\beta$-functions are calculated and display a surprising connection between classical and quantum physics: the classical integrability condition is necessary to prevent new couplings being generated by renormalisation.   We show these theories admit an elegant realisation of Poisson-Lie T-duality acting as a simple inversion of coupling constants.  The self-dual point corresponds to the Wess-Zumino-Witten model and is the IR fixed point under RG.   We address the possibility of having supersymmetric extensions of these models showing that extended supersymmetry is not possible in general.
}

\begin{document}

\pgfdeclarelayer{background layer} 
\pgfdeclarelayer{foreground layer} 
\pgfsetlayers{background layer,main,foreground layer}

\maketitle

\newpage

\section{Introduction}\label{s0} 
 
 Two-dimensional non-linear sigma models hold great interest for two key reasons.  First, they can provide prototypes with which to study strong coupling dynamics in a simpler setting than four-dimensional non-abelian gauge theories.   Second, they are the building blocks of the worldsheet description of string theory.   Under certain circumstances these theories can have a dramatic additional simplicity--that of integrability--allowing one to transcend the usual perturbative tool kit.   A rather long standing question has been to establish the complete landscape of integrable sigma models.  
 
 A substantial breakthrough was made by Klimcik with the explicit demonstration that the Yang-Baxter sigma models \cite{Klimcik:2002zj} are integrable \cite{Klimcik:2008eq}; thereby providing a one-parameter integrable deformation of the principal chiral theory associated to any semi-simple Lie algebra.  These theories, now often called $\eta$-deformations, have taken great prominence since they provide a Lagrangian description of a theory whose symmetry is deformed to a quantum group \cite{Delduc:2013fga}.  When extended to theories on symmetric spaces and to super-cosets, this has yielded a remarkable quantum group deformation of the $AdS_5 \times S^5$ superstring \cite{Delduc:2013qra} opening the door to an intriguing interpretation within holography.

 A surprising feature of the $\eta$-deformed theory in the context of the  $AdS_5 \times S^5$ superstring  is that it appears to describe a scale invariant but not Weyl invariant theory.  This is seen directly by the target spacetime's failure to satisfy the equations of Type IIB supergravity but instead to obey a set of ``generalised'' supergravity equations \cite{Arutyunov:2015mqj}.      Recent work has started to place these $\eta$-theories, and the generalised supergravity that govern their target spacetimes,   in the context of double/exceptional field theory \cite{Sakatani:2016fvh,Baguet:2016prz} and make explicit the link to T-folds and non-geometric configurations \cite{Fernandez-Melgarejo:2017oyu}. A  link between the  r-matrix, satisfying a (modified) classical Yang-Baxter equation,  that defines Yang-Baxter sigma models and the spacetime non-commutativity parameter  has been developed in \cite{Araujo:2017enj,Bakhmatov:2017joy} using the open-closed map.

 Notably, the $\eta$-theory displays a so-called Poisson-Lie (PL) symmetry. This means that it possesses a generalised T-dual in the Poisson-Lie sense proposed by Klimcik and Severa \cite{Klimcik:1995ux}. The Poisson-Lie dual model, modulo an analytic continuation, has been established to be a well-known integrable deformation called the $\lambda$-deformation.  Introduced by Sfetsos \cite{Sfetsos:2013wia} these theories interpolate between a Wess-Zumino-Witten (WZW)  \cite{Witten:1983ar} or a gauged WZW model  and the non-abelian T-dual of the principal chiral model on a group manifold or symmetric coset space respectively. The connection between the $\eta$- and the $\lambda$-theories was first shown for explicit $SU(2)$ based examples \cite{Hoare:2015gda,Sfetsos:2015nya} and established in generality by \cite{Klimcik:2015gba,Hoare:2017ukq}.  
     
 
Like the $\eta$-theories, $\lambda$-models can also be applied to cosets   \cite{Hollowood:2014rla} and semi-symmetric spaces 
 \cite{Hollowood:2014qma} and are thought to encapsulate quantum group deformations with $q$ a root of unity.  In contrast to the $\eta$-theory, the target spacetimes associated to the $\lambda$-model provide genuine solutions of supergravity (with no modification) \cite{Sfetsos:2014cea,Demulder:2015lva,Borsato:2016ose,Borsato:2016zcf,Chervonyi:2016ajp}.

 Given these successes a natural recent focus has been to understand potential generalisations of these approaches to include multi-parameter families of integrable models. On the side of the $\eta$-deformation (or Yang-Baxter model) notable are the two-parameter bi-Yang-Baxter deformations  \cite{Klimcik:2014bta}, the inclusion of a Wess-Zumino term \cite{Delduc:2014uaa} and indeed the recent synthesis of these \cite{Delduc:2017fib}. On the $\lambda$ side, multi-parameter deformations have been constructed and studied in \cite{Sfetsos:2014lla,Sfetsos:2015nya,Chervonyi:2016bfl,Appadu:2017bnv}.   There is also some evidence that a Poisson-Lie connection should be present between multi-parameter $\eta$- and $\lambda$-models; for example the  bi-Yang Baxter model has been shown to be related to a generalised $\lambda$-model \cite{Klimcik:2016rov}.  The Yang-Baxter theory with a WZ term (YB-WZ) appears amenable to similar treatment since it can be written as an ${\cal E}$-model \cite{Klimcik:2017ken} (though the corresponding $\lambda$ theory is not clearly spelt out as yet).  The construction of Lax pairs directly from the   ${\cal E}$-model  has recently been studied in  \cite{Severa:2017kcs}.

In this work we will provide further study of the multi-parameter YB-WZ model.  For the case of $SU(2)$ this system was studied in   \cite{Kawaguchi:2011mz,Kawaguchi:2013gma}. Specifically we shall,  

\begin{itemize}
\item Study the one-loop renormalisation of the general YB+WZ model extending results in the literature from $SU(2)$   \cite{Kawaguchi:2011mz} to arbitrary groups.  We will find that the conditions placed on a sigma model by integrability have an interesting interplay with renormalisation.  The condition required of classical integrability is preserved by RG flow.  Second, when dealing with non-simply laced algebras one finds the classical integrability condition is necessary for the renormalisation of the model not to introduce new couplings in addition to those of the bare theory.   That a classical property seems to be so tied to a very quantum calculation is notable.
\item We will clarify some details of the quantum group symmetries in these models and in particular show that the parameters defining the symmetry algebra are invariants of the RG flow. 
\item We comment on the role of  Poisson-Lie dualisation for the YB-WZ model.  Considered within the framework of the $\mathcal E$-model  \cite{Klimcik:2017ken}, the YB-WZ can be seen as being part of a pair of Poisson-Lie dual models. In particular, it admits a formulation as an $\mathcal E$-model associated to the Drinfeld double $\mathfrak d= \mathfrak g^{\mathbb C}$. When the integrability condition is satisfied,  the Poisson-Lie T-duality transformation preserves the structure of the action \eqref{eq:act} while the coupling parameters follow very simple ``radial inversion'' transformation rules. 

\item We will examine the possible worldsheet supersymmetrisation of the YB-WZ model associated to $SU(2)\times U(1)$ which is the simplest but non-trivial example that allows $N=(2,2)$ in the undeformed (WZW) case. While $N=(1,1)$ supersymmetry is always possible, going beyond that requires the introduction of additional geometric structures. We show that $N=(2,2)$ is forbidden for generic values of the deformation parameters while $N=(2,0)$ or $N=(2,1)$ is possible only for specific values. This leads us to conjecture that an $N=(2,2)$ YB-WZ model is not possible in general.
\end{itemize}

The paper is organised as follows. Section \ref{s1} introduces the Yang-Baxter Wess-Zumino model together with its integrability properties relevant to the subsequent discussions. In section \ref{s2} we give an explicit derivation of the one-loop $\beta$-functions of the YB-WZ model in the case of arbitrary groups. Given the result, we find that one needs to carefully distinguish between two cases: when the group is simply-laced or not. In the former case, a consistent renormalisation does not require the model to be integrable. For the latter case, the classical integrability condition turns out to be necessary to prevent the creation of new couplings in the theory by renormalisation. A detailed discussion of the RG behaviour is given in both cases. Section \ref{s4} formulates the YB-WZ action \eqref{eq:act} within the framework of the $\mathcal E$-model and derives the Poisson-Lie T-dual model. In section \ref{s6} we study the possibility of extended supersymmetry of the YB-WZ model. We end with a summary and conclusions in section \ref{s7}. The conventions used throughout this paper are given in appendix \ref{a0}. Appendix \ref{a1} reviews the construction \cite{Kawaguchi:2013gma} of the charges of the $SU(2)$ YB-WZ model  paying particular care to the overall normalisations required to expose the correct RG properties. In appendix \ref{s:properties} and \ref{s:geometry} we collate a set of useful expressions which were used in the calculations of the $\beta$-functions. 
 
     \section{Yang-Baxter and Yang-Baxter Wess-Zumino Models}\label{s1} 
       In this first section, we present the Yang-Baxter Wess-Zumino model (YB-WZ) as constructed in \cite{Delduc:2014uaa}, which will be the main topic of the remainder of this paper. Given a Lie algebra $\frak{g}$, we introduce an endomorphism $\cal{R}: \frak{g} \to    \frak{g}$ skew symmetric with respect to the Cartan-Killing product $\langle \cdot , \cdot  \rangle $  ($\langle {\cal{R}}x , y  \rangle=- \langle x, {\cal{R}} y  \rangle$) and obeying the modified classical Yang-Baxter (mCYBE) equation,
 \be \label{eq:mcybe}
[{\cal R} x,  {\cal R} y]  - {\cal R} \left( [x, {\cal R} y ] + [ {\cal R}x, y] \right)  = [x, y] \quad  \forall x, y \in \frak{g}  \ , 
 \ee 
 which further satisfies ${\cal R}^3= - {\cal R}$. The canonical realization of ${\cal R}$ is most easily seen in a Cartan-Weyl basis for the Lie algebra where it maps generators belonging to the CSA to zero and where it acts diagonally on generators corresponding to positive (negative) roots with eigenvalue $+i$ ($-i$).  Equipped with this structure, we define the YB-WZ action in worldsheet light-cone coordinates as,
 \begin{eqnarray}
\label{eq:act}
  &&{\cal S }= -\frac{1}{2\pi}\int d \sigma d\tau    \langle g^{-1} \partial_+g , \left(\alpha \mathbbm{1} + \beta {\cal R}  + \gamma  {\cal R}^2 \right)g^{-1} \partial_- g \rangle  \nonumber\\
&&\qquad\qquad + \frac{  k}{24\pi } \,      \int_{M_3}     \langle   \bar g^{-1} d\bar g, [\bar g^{-1} d\bar g,\bar g^{-1} d\bar g]  \rangle\,.
\end{eqnarray} 
Here as usual the coefficient of the Wess-Zumino term, $k$, is an integer, quantised such that the path integral based on this action is insensitive to the choice of the extension $\bar{g}: M_{3} \to G$. 
   
 A short calculation yields, after integration by parts and discarding the total derivative,  the equations of motion,
        \be
  \delta S = \frac{ \alpha- \gamma}{2\pi}\int d \sigma d\tau    \langle   \delta g g^{-1} , \partial_+{\cal K}_- + \partial_-{\cal K}_+  \rangle  \ ,
   \ee
  with, 
  \be
  {\cal K}_\pm = \frac{1}{\alpha - \gamma}  \left( (\alpha \mp  k)  \mp \beta {\cal R}_g + \gamma {\cal R}_g^2  \right) v_\pm  \ , \label{currents1}
  \ee
    in which we recall $v= dg g^{-1}$ are the right invariant one-forms and, 
    \be
     {\cal R}_g= \textrm{ad}_g \circ   {\cal R}  \circ \textrm{ad}_{g^{-1}}  \ , 
    \ee
    which, like ${\cal R}$, obeys the mCYBE and is skew symmetric with respect to the ad-invariant Cartan-Killing form $\langle \cdot , \cdot  \rangle$.   
    Using the inverse of eq.~(\ref{currents1}),
 \begin{eqnarray}
v_\pm&=&\left(\alpha-\gamma\right)\Big( \frac{1}{\alpha \mp k}\, \pm \,\frac{ \beta }{\beta^2+( \alpha \mp k- \gamma )^2}\,{\cal R}_g\nonumber\\
&& \qquad \qquad \quad +\frac{ \beta ^2- \gamma ( \alpha \mp k- \gamma )}{(\alpha \mp k)(\beta^2+( \alpha \mp k- \gamma )^2)} \,
{\cal R}_g^2\Big){\cal K}_\pm\,,
\end{eqnarray}    
in $dv-v\wedge v=0$, one easily gets,
\begin{eqnarray}
&&\partial _+{\cal K}_-- \partial _-{\cal K}_+- [{\cal K}_+,{\cal K}_-]= 
\Big(
 \frac k \alpha + \frac{\sqrt{ \gamma (\alpha ^2- \alpha \gamma -k^2)}}{\sqrt{\alpha }( \alpha-\gamma )}\,{\cal R}_g   \nonumber\\
 &&\hspace{5.9cm} -\frac{k \,\gamma }{\alpha (\alpha  -\gamma )}\,{\cal R}_g^2
\Big)\,( \partial_+{\cal K}_-+ \partial _-{\cal K}_+)\,, \label{osflat2}
\end{eqnarray} 
if and only if the coefficients are related via \cite{Delduc:2014uaa}, 
   \be
   \label{eq:intlocus}
   \beta^2 = \frac{\gamma}{\alpha} \left( \alpha^2 - \alpha \gamma - k^2 \right) \, .
   \ee 
So we conclude that the currents $ {\cal K}_\pm$ are on-shell flat  provided  eq.~(\ref{eq:intlocus}) holds. This is sufficient to guarantee classical integrability as the equations of motion follow then from the flatness of the standard $\frak{g}^{\mathbb{C}}$-valued Zakharov-Mikhailov Lax connection \cite{Zakharov:1973pp},
    \be\label{eq:lax}
\Lax_\pm(z) = \frac{1}{1 \mp z} {\cal K}_\pm  \ . 
   \ee

    We call the solutions to eq.~(\ref{eq:intlocus})  the {\em integrable locus}\footnote{To translate to    \cite{Delduc:2014uaa}   we have the dictionary of parameters $(\alpha, \beta, \gamma, k) \to (\eta^2,R, k', K)$ 
    \be
    A = \frac{\beta}{\alpha- \gamma} \ , \quad \eta^2 = \frac{\gamma}{\alpha  -\gamma} \ , \quad  k' = \frac{k}{\alpha - \gamma} \ , \quad  K= \frac{\alpha- \gamma}{4\pi} \ ,
    \ee
    however we shall continue with the $(\alpha, \beta, \gamma, k)$   such that $k$ gives the level of the WZW model that will appear at IR fixed points.  }. From eqs.~(\ref{osflat2}) and (\ref{currents1}) one deduces the further conditions $ \alpha \neq 0$ and $ \gamma \neq \alpha $, ensuring that the kinetic term is properly defined. In addition, as all parameters $\alpha $, $ \beta $, $\gamma$, $k$ are real and the kinetic term should have the right sign ($\alpha>0$), we conclude from eq.~\eqref{eq:intlocus} that the allowed values of $\alpha$ and $\gamma$ are $\alpha \in \left[ |k|\, ,\, \infty \right[$ and $\gamma \in \left[0\, ,\,\frac{\alpha^{2}-k^{2}}{\alpha} \right] $ or $ \alpha \in \left] 0\, ,\,|k|\right]$ and $\gamma \in \left[- \frac{k^{2}-\alpha^{2}}{\alpha}\, ,\,0 \right]$, where on $\alpha = |k|$ we find the WZW point \cite{Witten:1983ar}: $\gamma = \beta = 0$.
   
   For the particular subset of the integrable locus given by \cite{Klimcik:2002zj},
   \be
   k = 0 \ , \quad \alpha = \frac{1}{\tau} \ , \quad \beta = \frac{\eta}{\tau(1+\eta^2)} \ , \quad \gamma = \frac{\eta^2}{\tau(1+\eta^2) }  \ , 
   \ee
 the action eq.~\eqref{eq:act} reduces to what has become known as the $\eta$-deformed principal chiral model which is integrable \cite{Klimcik:2008eq} with the dynamics encoded in the flatness of a $\frak{g}^{\mathbb{C}}$-valued Lax connection $\Lax(z)$ depending on a spectral parameter $z\in \mathbb{C}$.  This theory displays a fascinating structure of infinite symmetries \cite{Delduc:2013fga}. At the Lagrangian level the left acting $G$ symmetry is preserved  and is complemented, as in the undeformed principal chiral model, with non-local charges furnishing a Yangian ${\cal Y}(\frak{g})$.  The right acting $G$ symmetry  is broken to its Cartan in the   action eq.~\eqref{eq:act},  but is enhanced by  non-local charges to form a classical version of a quantum group ${\cal U}_q( \frak{g})$  \cite{Delduc:2013fga} (actually   further extended to an affine ${\cal U}_q( \hat{ \frak{g}} )$   \cite{Delduc:2017brb}).  Schematically,   for a given simple root there exists a local charge $\frak{Q}^H $ and non-local charges $\frak{Q}^\pm$   that obey,
 \be
 \{ \frak{Q}^+, \frak{Q}^- \}  =   i \frac{q^{\frak{Q}^H} - q^{- \frak{Q}^H}  }{q -q^{-1} } \ , \quad  \{ \frak{Q}^\pm, \frak{Q}^H \}  =  \pm i \frak{Q}^\pm \ . 
 \ee 
 The quantum group parameter is given simply by $q= \exp( 8 \pi \tau \eta)$ which is an invariant under the renormalisation group flow of couplings \cite{Sfetsos:2015nya}. 
  
 The charges that generate these symmetries can be obtained by expansions around suitable values of the spectral parameter of the monodromy matrix,
  \be
U(z)= P\exp\left[ \int d\sigma \Lax_\sigma(z) \right] \ ,    
  \ee
 which is conserved by virtue of the flatness of $\Lax$.   The Yangian left acting symmetries are found through expansions around $z=  \infty $ whereas  the right acting quantum group symmetries are found  \cite{Delduc:2013fga}  via the expansion of the gauge transformed Lax around special points corresponding to poles in the twist function of the Maillet algebra \cite{Maillet:1985ek}.  
 
 Much of the story for the general $\eta$-deformed model was first established for the case of $\frak{g} = \frak{su}(2)$ which corresponds to the sigma model on a squashed $S^3$ (the Kalb-Ramond potential encoded by  eq.~\eqref{eq:act} is pure gauge in this case and though it doesn't effect the equation of motions it corresponds to an improvement term to ensuring flatness of currents).  The integrability was established many years ago by Cherednik  \cite{Cherednik:1981df}. Somewhat later the classical Yangian symmetry was shown in \cite{Kawaguchi:2010jg} and the (affine) quantum group symmetry in \cite{Kawaguchi:2011pf,Kawaguchi:2012ve}\footnote{  There is a small but potentially important subtlety here.  In  \cite{ Kawaguchi:2012ve} the affine charges are constructed from the expansion of a trigonometric Lax at infinity and appear in the principal gradation.  When the charges are extracted from the gauge transformation of the rational Lax evaluated around the poles in the twist function as in \cite{Delduc:2017brb} they appear in the homogeneous gradation; to go between the two gradations requires a spectral parameter dependent redefinition of generators. } .
  
Now we turn to the case where $k\neq 0$ which is the main focus of this paper.   Again historically this was first well explored for the case of $\frak{g} = \frak{su}(2)$.  The left acting symmetry is still  a Yangian \cite{Kawaguchi:2011mz} but the right acting symmetry is more mysterious \cite{Kawaguchi:2013gma} (we review the construction of the charges generating these generalised symmetries in appendix \ref{a1}).  One finds a structure similar to an  affine quantum group ${\cal U}_q( \widehat{ \frak{su}(2) })$ with\footnote{Here we restore the overall normalisations to the results in   \cite{Kawaguchi:2013gma} and map to our conventions.},
 \be
 q = \exp\left[ \frac{ 8 \pi \Theta}{\Theta^2 + k^2}\right] \ ,
 \ee 
but with a modification in how the affine tower of charges is build up. Namely, instead of taking successively the Poisson bracket to access the next charges in the tower, the Poisson bracket is multiplied at each step by an additional factor,
 \be
  \frac{\gamma^-}{\gamma^+} =  -\frac{k + i \Theta}{k- i \Theta} \ .
 \ee
 To move down, the Poisson bracket needs to be multiplied by its inverse (see figure 2 of \cite{Kawaguchi:2013gma} for further details). Here the combination,
\be
\Theta^2  = \frac{\alpha  \left(\alpha  (\alpha -\gamma )-k^2\right)}{\gamma } \ ,
\ee 
   will play a distinguished role in what follows; it will be seen to be an RG invariant.  
   
   As mentioned above, several partial results were already obtained for the YB-WZ model in the particular case where $\frak{g} = \frak{su}(2)$. In this paper we mostly focus on the case where $\frak{g}$ is arbitrary.  As we will see the general case shows several features which are absent when $\frak{g} = \frak{su}(2)$\footnote{Mathematically all differences between the general case and the simpler case where $\frak{g} = \frak{su}(2)$ arise from the fact that $\frak{su}(2)$ is the only simple Lie algebra where all roots are simple roots.}.

\section{Renormalisation of the YB-WZ model} \label{s2}
    
 Our aim is to calculate the $\beta$-functions for the couplings $\{ \alpha, \beta ,\gamma\}$ in the theory  defined by eq.~\eqref{eq:act}  without first assuming that the couplings lie on the integrable locus eq.~\eqref{eq:intlocus}.   The coupling $k$ being integer quantised evidently does not run.  To do so we will proceed geometrically; for a general two-dimensional non-linear sigma model the $\beta$-function for the metric $G_{\mu \nu}$ and Kalb-Ramond two-form potential   $B_{\mu \nu}$ in local coordinates $x^{\mu}$ are given by, at one-loop,  
 \be
  \begin{aligned}\label{eq:beta1}
 \mu \frac{d }{ d\mu} G_{\mu \nu} =   \hat\beta_{\mu \nu}^{G} &=  \alpha' \left(R_{\mu \nu} - \frac{1}{ 4} H^2_{\mu \nu}   \right)  + O( \alpha')^2 \ ,  \\ 
   \mu \frac{d }{ d\mu} B_{\mu \nu} = \hat\beta^{B}_{\mu \nu} &=    \alpha' \left( -\frac{1}{2} \nabla^\lambda H_{\lambda \mu \nu}  \right) + O( \alpha')^2 \ ,
    \end{aligned}       
   \ee
   where $H = \mathrm{d}B$ is the torsion 3-form and the connections and curvatures are to be calculated using $G$.  However, the diffeomorphism and gauge covariance of $G$ and $B$ means that these $\beta$-functions are ambiguous (even at one-loop order) 
 \cite{Shore:1986hk,Tseytlin:1986ws}  allowing us to modify them by\footnote{Note that the Lie derivative acts on $L_W B =  \iota_W H + d  \iota_W B$ and the latter term being a total derivative can be discarded within the action.},
  \be  \begin{aligned}\label{eq:diffeogauge}
\bar\beta^{G}_{\mu \nu } &=  \hat\beta^{G}_{\mu \nu }+ \nabla_{(\mu} W_{\nu)}\,, \\ 
\bar\beta^{B}_{\mu \nu }  &=     \hat\beta^{B}_{\mu \nu } + (\iota_W H)_{\mu \nu }  + (d\Lambda)_{\mu \nu}  \ , 
    \end{aligned} 
\ee
with $W$ and $\Lambda$ arbitrary target space one-forms. For the sigma model defined in eq. \eqref{eq:act}, of which the left acting $G$ symmetry is unaltered by the deformation, the target space data is most naturally expressed in a non-orthonormal frame formalism with frames defined by the left-invariant one-forms $u= g^{-1}dg  = - i u^A T_A$ as,
 \be\label{eq:metricansatz}
 G_{AB} =  \alpha\, \kappa_{AB} +\gamma\,  \R^{2}_{AB} \ ,
 \ee
 and torsion,
 \begin{eqnarray}\label{eq:torsion}
H_{ABC}= 3 \beta\, F_{\left[AB\right.}{}^{D}\R_{\left.C\right]D} - k\, F_{ABC} \ .
\end{eqnarray} 
Here $\kappa_{AB}  =  \langle T_A ,T_B \rangle$, ${\cal R}(T_A) = \R^{B}{}_{A} T_B$ and algebra indices out of position are lowered with $\kappa_{AB}$.  To completely fix things one should set $\alpha' =2$ so that the standard normalisation of the WZW models is recovered in the case  $\alpha = |k|, \beta =\gamma= 0$. 

 After a long battle making use of the properties listed in appendix \ref{s:properties} and the expressions of the geometry in the non-orthonormal frame listed in appendix \ref{s:geometry} one finds that the $\beta$-functions are given by,
 \begin{eqnarray}
  \hat\beta_{AB}^{G}  &= & -c_G\left(\frac{k^{2}(\alpha-2\gamma)}{2\alpha(\gamma-\alpha)^{2}} - \frac{\beta^{2}}{(\gamma - \alpha)^{2}} - \frac{\gamma^{2}}{2(\gamma - \alpha)^{2}} - \frac{1}{2} \right)\kappa_{AB}  -\frac{c_G}{2} \left(1-\frac{\alpha^{2}+\beta^{2}}{(\gamma - \alpha)^{2}}  \right) \R^{2}_{AB}\nonumber\\
&&  -\left( \frac{\gamma}{\alpha}\frac{k^{2}}{(\gamma-\alpha)^{2}} + \frac{\beta^{2}}{(\gamma - \alpha)^{2}} + \frac{\gamma}{(\gamma-\alpha)}  \right){\color{blue}F_{AD}{}^{C}F_{BC}{}^{E}\RR^{D}{}_E}\,,\label{eq:betaGSS}\\
  \hat\beta_{AB}^{B} & = &  \frac{2\beta}{\alpha-\gamma}{\color{blue} F_{AD}{}^{C} F_{BC}{}^E \R^{D}{}_{E}  } + c_G\,\frac{\beta}{\alpha-\gamma} {\cal R}_{AB}\,.\label{eq:betaBSS}
 \end{eqnarray}
The terms in blue involve tensor structures that are not present in the metric ansatz.  If these terms are not removed it would mean that under the RG flow the metric would flow out of the ansatz specified by eq.~\eqref{eq:metricansatz}.  Let us exploit the diffeomorphism symmetry to try and ameliorate the situation. With this in mind, note that for a one-form $W$ whose components $W_A$ are constant in frame indices we have:
 \be
 \begin{aligned}
\nabla_{\left(A\right.}W_{\left. B \right)} &= \frac{1}{2}\frac{\gamma}{\alpha - \gamma}\left( F_{AD}{}^{C}\RR^{D}{}_{B}+F_{BD}{}^{C}\RR^{D}{}_{A} \right)W_C\,,\\
(i_W H)_{AB} &=3\beta F_{\left[AB\right.}{}^{D}\R_{\left. C \right]D} G^{CE}W_E- k F_{ABC}G^{CD}W_D \,.
\end{aligned}
\ee
First, we try to use an appropriate choice of $W$ to remove the offending blue term in $\hat{\beta}^{G}$. However, using the properties listed in appendix \ref{s:properties}, one can show that the only sensible choice of $G^{-1}W$ involving the structure constants and the $\R$-matrix will always be Killing. Nevertheless, by taking the components $W_A$ proportional to $F_{AB}{}^{C} \R^{B}{}_C$, one can show that it is again Killing but can now in fact absorb the offending first term in $\hat{\beta}^{B}$. Finally, we remark that for the case of $\mathfrak{g} = \mathfrak{su}(2)$ ($c_G = 4$ in our conventions) the contribution of the parameter $\beta$ cancels exactly in $\hat{\beta}^{G}$ and can be gauged away by  an appropriate gauge choice $\Lambda$ in $\bar{\beta}^{B}$ eq.~\eqref{eq:diffeogauge} since $\R_{AB} u^A \wedge u^B$ is a pure gauge improvement term for $\mathfrak{su}(2)$.
      
We now consider the remaining offending term in $\hat{\beta}^{G}$ eq.~\eqref{eq:betaGSS}. Using a Cartan-Weyl basis for the Lie algebra and calling Lie algebra indices corresponding to positive (negative) roots as $a$, $b$, ... ($\bar a$, $\bar b$, ...) and those corresponding to directions in the CSA by $m$, $n$, ... one gets,
\begin{eqnarray}
F_{AD}{}^{C}F_{BC}{}^{E}\RR^{D}{}_{E}= c_G \kappa_{AB}+F_{Am}{}^CF_{BC}{}^m\,.
\end{eqnarray} 
The second term is non-vanishing only if the index $A$ corresponds to a positive root and the index $B$ to the corresponding negative root (or vice-versa) so one would expect it to be proportional to $ {\cal R}^2_{AB} $. Explicit computation gives,
\begin{eqnarray}
F_{am}{}^CF_{\bar a C}{}^m= -\kappa_{a\bar a}\,\vec a\cdot\vec a= \vec a\cdot\vec a\, {\cal R}^2_{a\bar a} \,,
\end{eqnarray} 
where $\vec a\cdot\vec a$ is the length squared of the root $a$. In our normalization it is always equal to 2 for simply laced groups ($\frak{g} =$ $A_n$, $D_n$, $E_6$, $E_7$ and $E_8$). For the non-simply laced groups its either 2 or 1 (for $\frak{g} =$  $B_n$, $C_n$ and $F_4$) or 2 or 1/3 (for $\frak{g} =$ $G_2$). So the term in blue in eq.~\eqref{eq:betaGSS} can be rewritten as,
\begin{eqnarray}
F_{AD}{}^{C}F_{BC}{}^{E}\RR^{D}{}_{E}= c_G \,\kappa_{AB}+2 \,x(A,B)\,{\cal R}^2_{AB} \,,
\end{eqnarray} 
where for simply laced groups $x(A,B)=1$ holds. For non-simply laced groups $x(A,B)$ assumes two different values pending the values of the indices $A$ and $B$. This implies that only for simply laced groups the  RG stays within the ansatz specified by eq.~\eqref{eq:metricansatz}. 

However, there is a second way to remain within the ansatz eq.~\eqref{eq:metricansatz}. Till now we did not impose any restriction on the parameters $ \alpha $, $\beta $, $ \gamma $ and $k$. Looking at the bothersome term in the last line of eq.~(\ref{eq:betaGSS}) we see that it precisely vanishes at the integrable locus eq.~(\ref{eq:intlocus}) and we remain within the ansatz eq.~\eqref{eq:metricansatz} for any group (simply laced and non-simply laced)! So we should distinguish two cases: case I, a restriction to the integrable locus for general groups, and case II, a restriction to simply laced groups where we can keep the parameters general. 

Before analysing both cases, we will consider a useful quantity to understand the RG flow: the Weyl anomaly coefficient $\tilde{\beta}^{\Phi}$. It is defined through the expectation value of the trace of the stress tensor,
 \be
 \langle T \rangle =   \frac{1}{ 4\pi } \tilde{\beta}^\Phi R^{(2)} + \dots \ , \quad 
\tilde{\beta}^\Phi = \frac{D}{6} -\alpha' \frac{ 1}{4} \left(R+4 \nabla^2 \Phi -4(\partial \Phi)^2 - \frac{1}{12}  H^2  \right) + O(\alpha')^2 \ . 
\ee
with $D$ the dimension of $\frak{g}$. The quantity $\tilde{\beta}^{\Phi}$, which one recognises in the spacetime effective Lagrangian for bosonic strings, can serve as a c-function for the models we are considering \cite{Tseytlin:1987bz}\footnote{In general one would need to average, i.e.\ integrate this over spacetime coordinates but the special form of the metric on a group manifold means that is not needed here.     }.  	Here we find for arbitrary groups in general,
   \begin{equation} 
\tilde{\beta}^\Phi = \frac{D}{6}+\frac{c_G D}{8} \left(\frac{1}{(\gamma-\alpha)} -\frac{k^{2}+\beta^{2}}{3(\gamma-\alpha)^{3}} \right) +\frac{c_G l}{8} \left( \frac{\gamma}{(\gamma - \alpha)^{2}} +\frac{k^{2}\gamma+\alpha \beta^{2}}{\alpha(\gamma-\alpha)^{3}} \right) \label{eq:dilbeta} 
\end{equation}
where $l$ is the rank of $\mathfrak{g}$. Focusing on the particular case of the integrable locus (i.e.\ case I) this equation reduces to,
\begin{equation}
\tilde{\beta}^\Phi= \frac{D}{6} + \frac{c_G D}{24}\left( \frac{(\Theta^{2}+\alpha^{2})(\alpha^{4} + k^{2}(\Theta^{2} - 3\alpha^{2}) - 3 \Theta^{2} \alpha^{2})}{\alpha^{3}(\Theta^{2} + k^{2})^{2}}  \right)\ .
\end{equation}
Whilst perhaps not so elegant, after applying the RG equations for this case we have,
  \be
\frac{d}{dt}   \tilde{\beta}^\Phi =\frac{c_G^{2} D \left(k^2-\alpha ^2\right)^2 \left(\alpha ^2+\Theta ^2\right)^2 \left(3 \alpha ^4-2 \alpha ^2 \Theta ^2+3
   \Theta ^4\right) }{48\alpha ^6 \left(\Theta
   ^2+k^2\right)^4} \ . 
 \ee
Notice that because $\left(3
   \alpha ^4+3 \Theta^4-2 \alpha ^2 \Theta^2\right)$ has no real roots for  $\alpha^2 \in \mathbb{R}$ and $\Theta^2 \in \mathbb{R}$ we explicitly see the monotonicity of the flow $ 
    d_t   \tilde{\beta}^\Phi  > 0 
 $ with $t\rightarrow \infty$ in the UV giving as required $  \tilde{\beta}^\Phi|_{UV} >  \tilde{\beta}^\Phi|_{IR}$.   The IR is no more than the WZW CFT at $\alpha= |k|, \gamma=\beta=0$ (for which of course  $ \tilde{\beta}^\Phi  = \frac{D}{6} - \frac{D h^v}{6 k }  + O(\frac{1}{k})^2$  in accordance with the large level expansion of $\frac{1}{6}$ times the central charge  $ c= \frac{k \dim G}{ k+ h^v }$).

\subsection{Case I: general group $G$ and restriction to the integrable locus}

\noindent We will now restrict ourselves to the integrable locus, i.e.\ the coupling constant $\beta $ satisfies eq.~(\ref{eq:intlocus}), 
\begin{eqnarray}
\beta= \pm \sqrt{\frac{\gamma }{\alpha }}\, \sqrt{ \alpha ^2- \alpha \gamma -k^2}\, ,
\end{eqnarray} 
whilst keeping the group $G$ arbitrary. Eqs. (\ref{eq:betaGSS}) and (\ref{eq:betaBSS}) now become,
 \begin{eqnarray}
  \hat\beta_{AB}^{G}  &= & -c_G\left(\frac{k^{2}(\alpha-2\gamma)}{2\alpha(\gamma-\alpha)^{2}} - \frac{\beta^{2}}{(\gamma - \alpha)^{2}} - \frac{\gamma^{2}}{2(\gamma - \alpha)^{2}} - \frac{1}{2} \right)\kappa_{AB}  -\frac{c_G}{2}  \left(1-\frac{\alpha^{2}+\beta^{2}}{(\gamma - \alpha)^{2}}  \right) \R^{2}_{AB}\nonumber\\
&=& - \frac{c_G}{2}\, \frac{k^2- \alpha ^2}{( \alpha- \gamma )^2}\,\kappa_{AB}-
\frac{c_G}{2}\, \frac{ \gamma }{\alpha ( \alpha- \gamma )^2}\, \Big(2 \alpha \gamma -3 \alpha ^2+k^2\Big){\cal R}^2_{AB} 
\,,\label{eq:betaGSSint}\\
  \hat\beta_{AB}^{B} & = &  c_G\,\frac{\beta}{\alpha-\gamma} {\cal R}_{AB} \nonumber\\
&=&\pm c_G \,\frac{1}{\alpha-\gamma}\, \sqrt{\frac{\gamma }{\alpha }}\, \sqrt{ \alpha ^2- \alpha \gamma -k^2}\, {\cal R}_{AB}
\,.\label{eq:betaBSSint}
 \end{eqnarray}
 \noindent Eq.~(\ref{eq:betaGSSint}) yields the RG equations for the independent coupling constants $\alpha $ and $\gamma $,
 \begin{eqnarray}
\frac{d \alpha}{dt} &=& - \frac{c_G}{2}\, \frac{k^2- \alpha ^2}{( \alpha- \gamma )^2}\,,\nonumber\\
\frac{d \gamma }{dt} &=&-\frac{c_G}{2}\, \frac{ \gamma }{\alpha ( \alpha- \gamma )^2}\, \Big(2 \alpha \gamma -3 \alpha ^2+k^2\Big)\,.\label{RGinte}
\end{eqnarray} 
Note that eq.~(\ref{RGinte}) is simply a rescaling of that obtained for $\frak{su}(2)$ in \cite{Kawaguchi:2011mz}. Therefore, the group dependence in the flow equations is limited to the rate of the flow. Indeed, absorbing the factor $c_G$ in the RG time, $t\rightarrow c_G t$, the flow can be made independent of the Lie group $G$.
Eq.~(\ref{eq:betaBSSint}) also consistently yields the flow of the dependent parameter $\beta $,
\begin{eqnarray}
\frac{d \beta}{d t} = c_G\,\frac{\beta}{\alpha-\gamma}\,.
\end{eqnarray} 
Using these equations one immediately gets,
\begin{eqnarray}
\frac{d}{d t} \left( \beta^2 - \frac{\gamma}{\alpha} \left( \alpha^2 - \alpha \gamma - k^2 \right)  \right) = \frac{2 c_G}{\alpha- \gamma} \left( \beta^2 - \frac{\gamma}{\alpha} \left( \alpha^2 - \alpha \gamma - k^2 \right)  \right)  \,, 
\end{eqnarray} 
showing that the integrable locus is preserved by the RG! 

\noindent Moreover, this system has an  RG invariant aside from the coefficient  of the WZ term,
      \be \label{eq:RGinv}
\Theta^2 = \frac{\alpha  \left(\alpha ^2-\alpha  \gamma -k^2\right)}{\gamma } \ ,
      \ee
 in terms of which we have a single independent RG equation,  
 \be
\frac{d}{dt}\alpha =   \frac{c_G}{2} \frac{ ( \alpha^2 -k^2 )   \left(\alpha ^2+\Theta^2\right)^2}{\alpha ^2
   \left(k^2+\Theta^2\right)^2} \ .
   \ee
Returning to the discussion in section \ref{s1} we see that the parameters entering the charge algebra are RG invariants since they are functions of $\Theta$ and $k$ alone.

\subsubsection*{Discussion of the RG behaviour at  the integrable locus} 

\begin{figure}[H]
\centering
\includegraphics[scale=0.8]{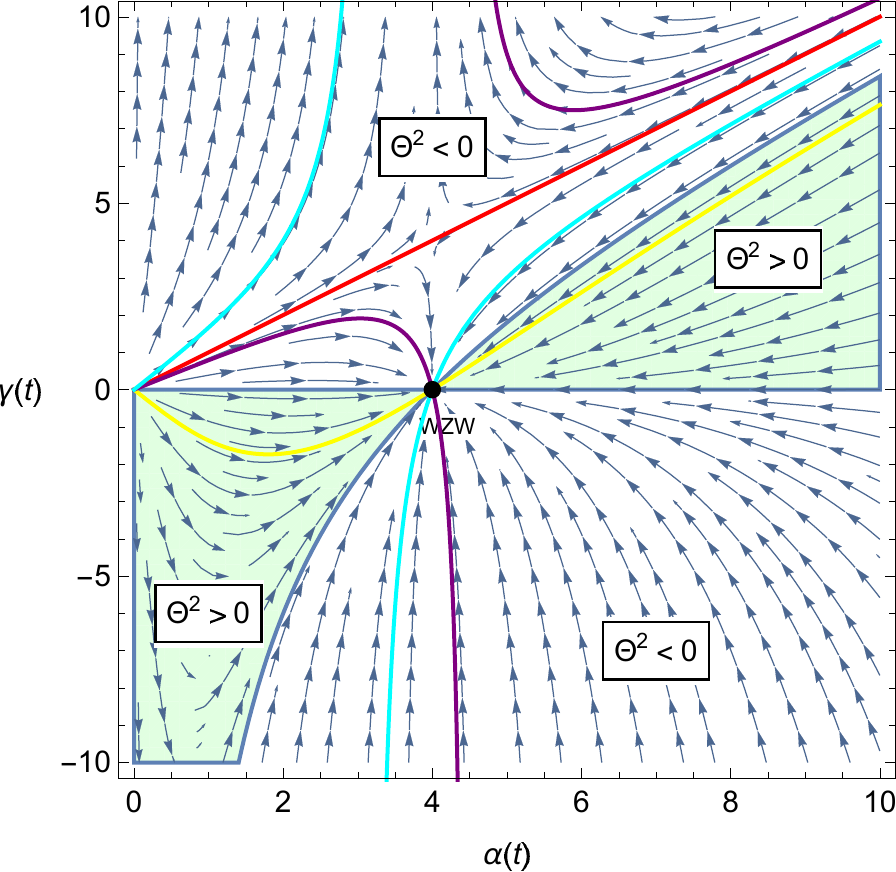}
\caption{The RG evolution for the $G=SU(3)$ and $k=4$ YB-WZ model restricted on the integrable locus of the couplings $\alpha$ vs.\ $\gamma$. The arrows point towards the IR. The red line $\alpha=\gamma$ depicts the points where the one-loop result blows up and it is labeled by $\Theta^{2}=-k^{2}=-16$. The black dot represents the WZW point ($\alpha=4$, $\beta=\gamma=0$) which is an IR fixed point with only irrelevant directions. The green regions are those where the coupling $\beta(\alpha,\gamma)$ and the RG invariant $\Theta$ are real. The yellow line portrays $\Theta^{2} = 10$, the cyan line $\Theta^{2}=-10$ an the purple line $\Theta^{2}=-20$.}\label{fig:RGint}
\end{figure}

The case of $SU(2)$ was already considered in \cite{Kawaguchi:2011mz}, where at first sight it appears to be different because the $\beta$ coupling is a total derivative in the Lagrangian and serves merely as an improvement term in the currents. The renormalisation of this coupling in the case of $SU(2)$ can be absorbed by a gauge transformation generated by $\Lambda$ of eq.~\eqref{eq:diffeogauge}. So in fact the analysis of the RG phase portraits performed in \cite{Kawaguchi:2011mz} is equally valid here, corroborating the group dependence of the flow. However, for completeness and later discussion we present in figure \ref{fig:RGint} the RG behaviour of the $G=SU(3)$ YB-WZ model at level $k=4$ restricted to the integrable locus. 

In this case, we have an RG invariant $\Theta$ given by eq.\ \eqref{eq:RGinv} which labels the RG trajectories. The only fixed point is now the WZW at $\alpha=|k| =4$, $\beta=\gamma=0$ in the IR. Again, on the $\alpha = \gamma$ line the one-loop result blows up and the metric is degenerate. Since we are restricted to the integrable locus, where $\beta$ satisfies eq.\ \eqref{eq:intlocus}, the physically allowed theories are located in the regions where $\beta$, or equivalently the RG invariant $\Theta$, is real. There are two such regions indicated in green.
  
\noindent A physically allowed trajectory is portrayed by the yellow line in figure \ref{fig:RGint}, along which the RG invariant has the constant real value. By varying the value of the RG invariant $\Theta \in \mathbb{R}$, we can cover the full region of physically allowed trajectories. In the green region where $\gamma<0$, we start from the trivial fixed point at $\alpha=\gamma=0$ in the UV and end up at the WZW in the IR in a finite RG time. In the green region where $\gamma>0$, the WZW is again an IR fixed point but the asymptotic behaviour is not yet apparent.  However, as we will see in the coming section, these two green regions are to be physically identified by a duality.

\noindent  In the other regions, we have either $0>\Theta^{2} > -k^{2}$, represented by the cyan line, or $\Theta^{2}<-k^{2}$, represented by the purple line.  The crossover is given by $\Theta^{2}=-k^{2}$ which corresponds to the red $\alpha=\gamma$ line. In any case, we flow either to the WZW or to a {\color{black}strongly $\gamma$-coupled theory} in the IR. Conversely, flowing towards the UV leads either to the trivial fixed point or to an {\color{black}unsafe} theory.

Let us analyse the behaviour around the IR WZW fixed point.   If we linearise the flow around the fixed point, i.e. let $\alpha = k + \bar\alpha$ and $\beta =0+ \bar\beta$ $\gamma = 0+ \bar \gamma$, we see from   eqs.~\eqref{RGinte}, 
     \begin{eqnarray}
  \label{eq:RGEs2 } 
\frac{d \bar \alpha}{dt} =  \frac{c_G}{k }\, \bar \alpha  \,,\quad
 \frac{d \bar \beta}{dt} =  \frac{c_G}{k }\, \bar \beta  \,,\quad
 \frac{d \bar \gamma }{dt} =  \frac{c_G}{k }\, \bar \gamma  \, .
 \end{eqnarray}  
 Since they all have positive sign's on the right-hand side we conclude that these are indeed irrelevant. Making use of the RG invariant eq.~\eqref{eq:RGinv} and   the integrable locus eq.~\eqref{eq:intlocus} we can express the action as,
  \be 
\label{eq:act2}
 {\cal S }= -\frac{1}{2\pi}\int d \sigma d\tau    \Tr\left(  g^{-1} \partial_+g , \left(\alpha \mathbbm{1} + \frac{\Theta (\alpha^2 - k^2) }{\alpha^2 + \Theta^2} {\cal R}  +  \frac{\alpha (\alpha^2 - k^2) }{\alpha^2 + \Theta^2}    {\cal R}^2 \right)g^{-1} \partial_- g \right) +  I_{WZ}\ ,
  \ee
 where we choose the positive sign for the $\beta$-coupling. Now expanding around the IR fixed point to leading order in $\bar{\alpha}$  we have, 
    \be 
\label{eq:act3}
 {\cal S }=  S_{WZW_k}[g] -\frac{\bar{\alpha}  }{2\pi}\int d \sigma d\tau     \Tr\left(  g^{-1} \partial_+g , \left(  \mathbbm{1} + \frac{2 k\Theta  }{k^2 + \Theta^2} {\cal R}  +  \frac{ 2 k^2  }{k^2 + \Theta^2}    {\cal R}^2 \right)g^{-1} \partial_- g \right)  \ .
   \ee
  To interpret this let us now go to the Euclidean setting and define the usual WZW CFT currents,   
  \be
  J(z)= J^a(z) t^a = - \frac{k}{2} \partial g g^{-1} \ , \quad   \bar J(\bar z) =   \frac{ k}{2}    g^{-1} \bar\partial g   \ , \quad   
  \ee   
   which obey  a current algebra,
    and are Virasoro primary with weights  $(1,0)$ and $(0,1)$ with respect to the  Sugawara stress tensor. 
   Consider a composite field $\phi_{\ell \bar{\ell} }(z, \bar{z})$   transforming in representations labelled by $\ell$ and $\bar{\ell}$ under the affine $G_L \times G_R$ symmetry.  This field will also be Virasoro primary and will be have an anomalous dimensions $(\Delta_\ell, \bar\Delta_{\bar \ell}) $.   As explained in \cite{Knizhnik:1984nr} the associated representation of the full Virasoro $\rtimes$ KM algebra is degenerate with a null vector.  Because of this the anomalous dimension can be extracted as, 
   \be
   \Delta_\ell = \frac{c_\ell}{c_G + k} \ ,
   \ee
   where $c_\ell \mathbb{I} = t^a_\ell t^a_\ell$.   Examples of such primaries are $g(z,\bar z)$, the group element itself, but also composites including the adjoint action,
      \be
  {\cal D}^{ab}(z,\bar{z})  = tr(g^{-1} t^a g t^b )  \ , 
  \ee
  that transforms in the adjoint of $G_L$ on the first index and the adjoint of $G_R$ on the second.  This operator  has anomalous dimension $\Delta_{{\cal D}}  =  \bar\Delta_{{\cal D}}  = \frac{c_G}{c_G+ k}$  and   can be used to define the ``wrong'' currents i.e., 
\be
K = K^a t^a =  - \frac{k}{2}g^{-1} \partial g   = g^{-1} J g \Rightarrow K^a = J^b {\cal D}^{b a}  \ ,
\ee
 with dimensions $(1+ \Delta_\ell, \bar\Delta_{\bar \ell}) $.

 Now we can see that the deforming operator is of the form, 
\be\label{eq:Defop}
{\cal O}(z,\bar{z}) \sim   K^a(z,\bar{z}) {\cal M}_{ab} \bar{J}^b(\bar z)   \ , \quad  {\cal M} = \left(  \mathbbm{1} + \frac{2 k\Theta  }{k^2 + \Theta^2} {\cal R}  +  \frac{ 2 k^2  }{k^2 + \Theta^2}    {\cal R}^2 \right)\ ,
\ee
 and has total dimension $2 +2 \Delta_{\cal D} > 2$ and is irrelevant even without any further corrections.   Suppose that we send $\Theta \to \infty$ then we are in exactly the situation considered in \cite{Witten:1983ar,Knizhnik:1984nr} of the flow of the PCM plus  a Wess-Zumino term with the WZW as the IR fixed point.    
 
 Now recall that the Callan-Symanzik equation can be used to relate the beta function to the anomalous dimension and indeed we see that in the large $k$ limit (in which loop corrections are suppressed) the anomalous dimension of ${\cal O}$, $\gamma_{\cal O}  \to \frac{c_G}{k}$ precisely in agreement with the leading order of the beta functions  eq.~\eqref{eq:RGEs2 }.
 
 It would be interesting to develop this line further and to try and ascertain all loop summation of the anomalous dimension following similar techniques to those adopted in the context of $\lambda$-models in   \cite{Georgiou:2015nka}.  There is however an added complexity that the deforming operator is not diagonal in the algebra indices but mixed with the inclusion of the ${\cal M}$ matrix.

\subsection{Case II: simply laced groups and general parameters}

Although it is outside the primary purpose of this paper -- which is to study integrable deformations -- it is intriguing to look at the case of simply laced groups for which a consistent renormalisation did not require the model to lie on the integrable locus.  It is then possible to rewrite eqs. (\ref{eq:betaGSS}) and (\ref{eq:betaBSS}) as,
\begin{eqnarray}
\hat\beta_{AB}^{G}  &=& -\frac{c_G}{2} \,\frac{k^2- \alpha ^2}{( \alpha- \gamma )^2}\, \kappa_{AB} 
 -\Bigg[\frac{c_G}{2} \left(1-\frac{\alpha^{2}+\beta^{2}}{(\gamma - \alpha)^{2}}  \right)  +
2\left( \frac{\gamma k^{2}+\alpha \beta^{2}}{\alpha(\gamma-\alpha)^{2}} + \frac{\gamma}{(\gamma-\alpha)}  \right) \Bigg]\R^{2}_{AB}\,, \nonumber
\label{eq:betaGSSsl}\\
  \hat\beta_{AB}^{B} &=&  c_G\,\frac{\beta}{\alpha-\gamma} \R_{AB}\,.\label{eq:betaBSSsl}
 \end{eqnarray}
 This gives the following RG equations for the coupling constants $\alpha $, $\beta $ and $ \gamma $ (with RG time $t =\log\mu$):
 \begin{eqnarray}\label{eq:RGsl}
\frac{d \alpha }{dt}&=&  -\frac{c_G}{2} \,\frac{k^2- \alpha ^2}{( \alpha- \gamma )^2}\,,\nonumber\\
\frac{d \beta  }{dt}&=&  c_G\,\frac{\beta}{\alpha-\gamma}\,,\nonumber\\
\frac{d \gamma  }{dt}&=&- \frac{c_G}{2} \left(1-\frac{\alpha^{2}+\beta^{2}}{(\gamma - \alpha)^{2}}  \right)  -2\left(\frac{\gamma k^{2}+\alpha \beta^{2}}{\alpha(\gamma-\alpha)^{2}} + \frac{\gamma}{(\gamma-\alpha)}  \right)\,.
\end{eqnarray} 
We will analyse the RG behaviour in some detail below. However one already notices a remarkable fact. Besides the standard WZW fixed point ($ \alpha =|k|$, $ \beta =\gamma =0$), a second fixed point seems to emerge at $\alpha =|k|$, $ \beta =0$ and $ \gamma =2c_G|k|/(c_G+4)$ iff.\ $c_G \neq 4$ or thus $G \neq SU(2)$.  We call this point FP2.    When $G=SU(2)$ the RG equations blow up on the FP2 values (since then $\alpha=\gamma$) and the second fixed point is removed. Furthermore, for $SU(2)$ one sees that the terms involving $\beta$ cancel in the flow equation for $\dot\gamma$  and the general RG equations of the remaining $\alpha$ and $\gamma$ will coincide with the corresponding RG equations when restricted to the integrable locus (see above).

\subsubsection*{The RG behaviour when not restricted to the integrable locus}

\begin{figure}[H]
	\centering
	\begin{subfigure}[t]{0.49\textwidth}
		\centering
		\includegraphics[width=0.97\textwidth]{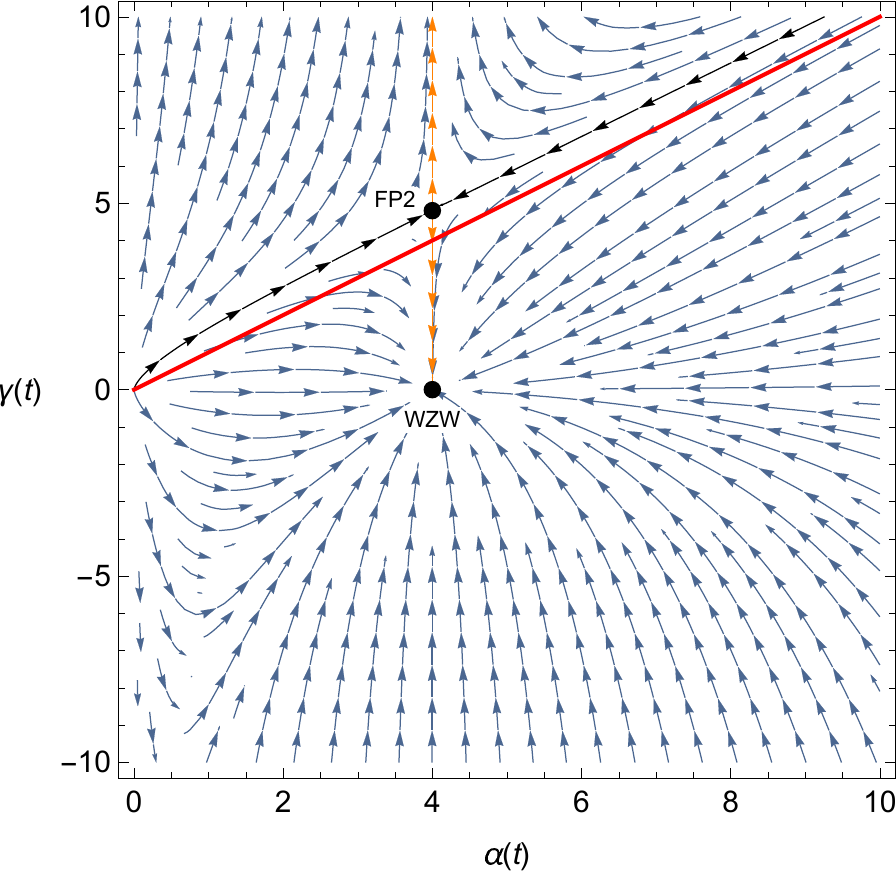}
		\caption{ The RG flow in the $\beta=0$ slice. }\label{fig:RGslBF}		
	\end{subfigure}
	\begin{subfigure}[t]{0.49\textwidth}
		\centering
		\includegraphics[width=0.97\textwidth]{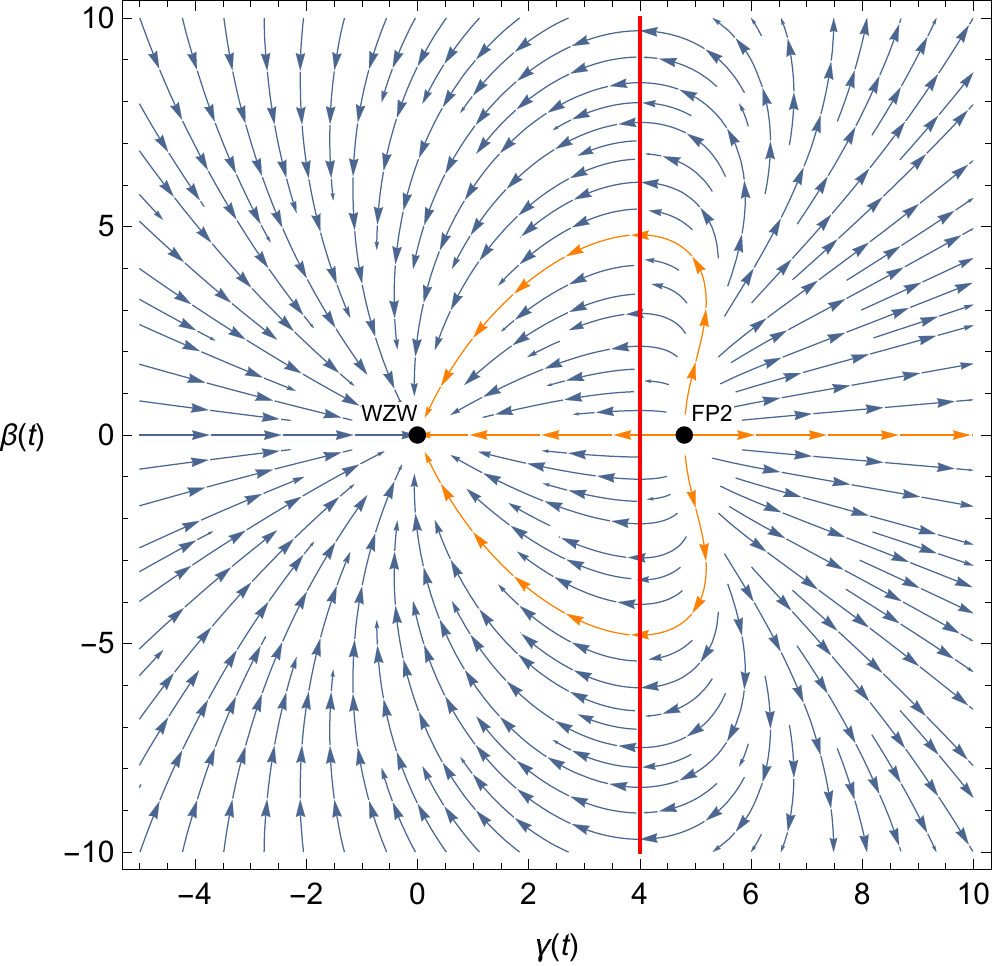}
		\caption{The RG flow in the $\alpha=4$ slice.}\label{fig:RGslAF}
	\end{subfigure}
	\caption{The RG evolution of $(\alpha,\beta,\gamma)$ for $G=SU(3)$ and $k=4$. The arrows point towards the IR. The red line $\alpha = \gamma$ depicts the points where the one-loop result blows up. The black dots represent the RG fixed points WZW and FP2. We see that FP2 exhibits two relevant (orange lines) and one irrelevant (black line) direction. The WZW is a true IR fixed point with only irrelevant directions. }\label{fig:RGsl}
\end{figure}

To illustrate the existence of the second fixed point FP2, we consider the RG flow for the case of  the group $G=SU(3)$, setting $k=4$. We plot the flow in two slices of the three-dimensional coupling space $(\alpha,\beta,\gamma)$ in order to visualise various directions around the fixed points. Figure \ref{fig:RGslBF} shows the flow of $\alpha$ vs.\ $\gamma$ in the $\beta=0$ slice and figure \ref{fig:RGslAF} the flow of $\gamma$ vs.\ $\beta$ in the $\alpha=4$ slice. \\


%

From the above figures \ref{fig:RGslBF} and \ref{fig:RGslAF}, we see qualitatively that the WZW fixed point exhibits three independent irrelevant directions and the FP2 fixed point one irrelevant and two relevant independent directions. This can be made precise by again analysing the linearlised flows in the neighbourhood of the fixed points. In a more compact notation, denoting $\xi_i = \{\alpha,\beta,\gamma\}$ with $i\in \{1,2,3\}$, the linearlised flow can be written as,
\begin{equation}
\frac{\partial \xi_{i}}{\partial\log\mu} = A_{ij}\delta\xi_{i} + \mathcal{O} (\delta\xi_{i}^{2})\,.
\end{equation}
In the neighbourhood of the WZW point we find,
\begin{equation}
A^{\text{WZW}}_{ij} =   \frac{c_G}{|k|}\delta_{ij}\,,
\end{equation}
which gives indeed three independent irrelevant directions (they have positive eigenvalues). On the other hand, in the neighbourhood of the second fixed point we find,
\begin{equation}
A^{FP2}_{ij} = \frac{c_G(c_G+4)}{|k|(c_G-4)}\begin{pmatrix}
\frac{c_G +4}{c_G - 4} &0&0\\
0&-1&0\\
\frac{2(c_G+4)}{c_G-4}&0&-\frac{c_G+4}{c_G-4}
\end{pmatrix}\,,
\end{equation}
for which the eigenvalues read:
\begin{equation}
 \left\{\frac{c_G(c_G +4)^{2}}{|k|(c_G-4)^{2}}, -\frac{c_G(c_G+4)}{|k|(c_G-4)}, -\frac{c_G(c_G+4)^{2}}{|k| (c_G-4)^{2}}   \right\}.
\end{equation}
Thus, from the second fixed point indeed two relevant and one irrelevant independent directions emerge.


At the two fixed points  the Ricci curvature $R$ \eqref{eq:riccicurv} evaluates to,
\begin{align}
R_{\text{WZW}} &= \frac{c_G D}{4|k|}, \nonumber\\
R_{\text{FP2}} &= - \frac{c_G }{4 (c_G -4)^2}\left(\left(c_G^2 - 16 \right) D + 2 c_G \left( c_G +4 \right) l \right)       \frac{1}{|k|},\nonumber
\end{align}
with $D$ the dimension and $l$ the rank of $G$ such that the target spaces are weakly curved for large enough $k$ and for which the one-loop result is trustworthy.   Whilst there is no reason to believe that the location of FP2 (i.e.\ the value of $\gamma$ at the fixed point) is one-loop exact, it seems likely that its existence is robust to loop corrections.  It is then conceivable that FP2 may define a CFT.  This being the case, from the general dilaton $\beta$-function eq.~\eqref{eq:dilbeta} we can read off the effective central charge $c_\text{eff} $ of FP2 at one-loop  to find,
\begin{equation}
c_\text{eff} = D + \frac{c_G (c_G +4) \left( (16 + (c_G -16)c_G)D + 6 c_G^{2} l \right)}{2 |k| (c_G - 4)^{3}}\,.
\end{equation}
Before further discussing this possibility let us explore other aspects of the RG flow. 

\begin{figure}[H]
\centering
\includegraphics[scale=0.7]{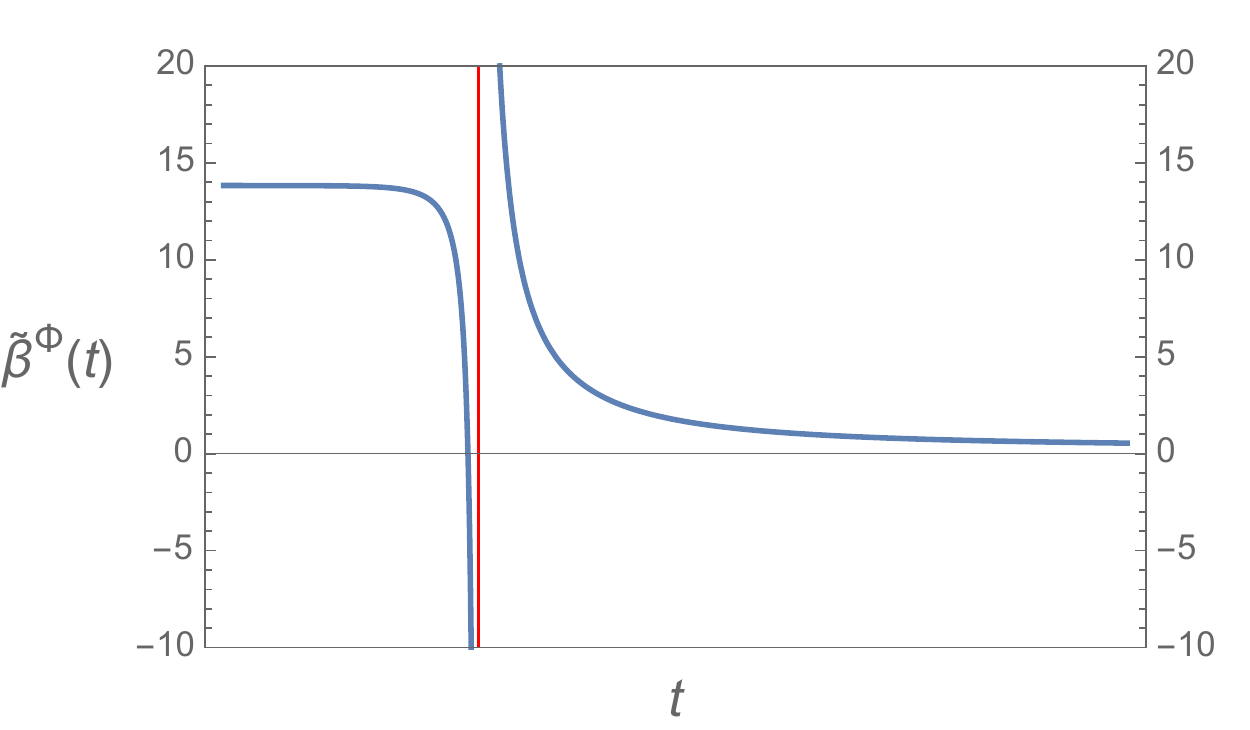}
\caption{The Weyl anomaly coefficient $\tilde{\beta}^\Phi$ evaluated for $G=SU(3)$ and $k=4$ along the orange RG trajectory $\beta=0, \alpha=|k|$ shown in figures \ref{fig:RGslBF} and \ref{fig:RGslAF} that connects the UV FP2 to the WZW in the IR. } \label{fig:BetaPhi}
\end{figure}

At first sight, and consistent with $c_{UV} > c_{IR}$, is that FP2  defines a UV fixed point from which in the deep IR one arrives at the WZW theory.   However, care has to be taken when traveling over the line $\alpha=\gamma$, displayed in red in the figures. In the vicinity of this line the one-loop approximation is evidently not trustworthy; the target space curvatures blow up  for small values of $|\gamma-\alpha|$ as is clear from the curvature $R$ eq.~\eqref{eq:riccicurv} and indeed the metric   $G_{AB}$ \eqref{eq:metricansatz} becomes degenerate.   In light of the apparent singularity of the one-loop flow equations where $\gamma-\alpha$ appears in denominators,  it is then quite surprising that numerically a global picture emerges with flows that transgress the red line.  We are then led to ask if such an RG trajectory can cross the $\alpha=\gamma$ line in a finite RG time. To show that this is possible  we concentrate on the slice of $\alpha= |k|$ illustrated in fig.~\ref{fig:RGslAF} and for further simplicity consider going backwards along the orange direction $\beta=0$ starting near to the WZW point.  Along this trajectory we can calculate the RG time $\Delta t$, with $t= \log \mu$, by evaluating,
\begin{eqnarray}
\Delta t &=& \int_{\gamma=\gamma_i}^{\gamma= \gamma_f} \frac{d t }{d\gamma} d\gamma  \, .
\end{eqnarray} 
One can show that there is no pathology associated with $\gamma= \alpha =|k|$ in this quantity.     Given this, one is encouraged to take seriously the quantity $\tilde{\beta}^\Phi$ defined in eq.~\eqref{eq:dilbeta}   as a would-be c-function for the flow connecting FP2 and WZW.  For simplicity we again consider this quantity along the orange direction $\beta=0,\alpha=|k|$   in fig.~\ref{fig:RGslAF} and plot the result in figure  \ref{fig:BetaPhi}.  What we see is that $\tilde{\beta}^\Phi$ is sensitive, unsurprisingly, to the singularity at $\gamma = k$.  Whilst $\tilde{\beta}^\Phi|_{UV}>  \tilde{\beta}^\Phi|_{IR}$ and its derivative is strictly positive, it is not a positive definite quantity and diverges at    $\gamma = k$.  Of course one should not read too much into this; the singularity is just symptomatic of the breakdown of the perturbative approximation.  One could still expect that a correct  strictly positive monotonic function exists and it agrees with this one-loop approximate result where the one-loop result is valid.  

   Combining the observation that  the one-loop approximation is robust around the fixed points (for large $k$) and the unexpected global continuity of the numerical solutions in figs.~\ref{fig:RGslBF} and \ref{fig:RGslAF} leads us to tentatively suggest that there is indeed an RG flow between a new fixed point FP2   and an IR WZW model but that the sigma model description may not be the correct variables to reveal this.    
          
  There are several points that  merit investigation:
   \begin{itemize}
   \item At FP2, $\det(G_{AB}) < 0$.  This means that some currents occur in the action with a negative coefficient of their kinetic term. A conservative viewpoint would be to regard this as non-physical but this then begs the  question what is the UV completion of the model? Let us instead take FP2 seriously. Should FP2 define a CFT, it is presumably non-unitary.   In this case we would have an RG flow between a non-unitary UV theory and an unitary IR theory. Perhaps this suggestion is not as outlandish as might first seem.   By way of example we could consider RG flows in minimal models.  It is well known \cite{Ludwig:1987gs,Zamolodchikov:1987ti}  that the $\phi_{(1,3)}$ deformation of the $p^{th}$ minimal model ${\cal M}_{(p,p+1)}$ triggers an RG flow resulting in the $(p-1)^{th}$ minimal model IR\footnote{More precisely this occurs when the deformation parameter is negative, when the deformation parameter is positive the flow results in a massive theory.}. Less familiar perhaps are the RG flows involving non-unitary minimal models, i.e.\  ${\cal M}_{(p,q)} $ with  $p,q$ co-prime and $q \neq p+1$, whose study was initiated in \cite{Lassig:1991an,Ahn:1992qi,Martins:1992ht}.  More generally \cite{Dorey:2000zb}, chains of non-unitary minimal models can be connected by RG flows triggered by alternating deformations of $\phi_{(1,5)}$ and $\phi_{(2,1)}$ which then terminate in an unitary minimal model.   One example from \cite{Dorey:2000zb} is\footnote{Notice that this terminates in the unitary critical Ising model with $c=\frac{1}{2}$ and, just as with the flow between tri-critical Ising and  critical Ising, the single massless Majorana fermion of the final IR theory can be interpreted as the goldstino for the spontaneous breaking of the supersymmetry present in ${\cal M}_{(3,8)}$. }, 
   \be
 \dots  {\cal M}_{(5,12)} \xrightarrow{  \phi_{(1,5)} } {\cal M}_{(5,8)} \xrightarrow{  \phi_{(2,1)} }{\cal M}_{(3,8)}  \xrightarrow{  \phi_{(1,5)} }{\cal M}_{(3,4)}  \ .
   \ee
%
%
An other example in  \cite{Dorey:2000zb} terminates in 
   a flow from the Yang-Lee edge singularity to the trivial $c=0$ theory. Recently in  \cite{Castro-Alvaredo:2017udm} it was shown that it is possible to relax the requirement of unitary and still show the existence of a monotonic decreasing c-function along such flows.  So the learning here is that it is not a manifest impossibility to conceive an RG flow between a non-unitary UV CFT and a unitary IR CFT. 
   
   \item The fate of FP2 with regard to higher loop corrections needs to be established; does it persist? 
   \item Is the postulated  FP2 both scale {\em and } Weyl invariant\footnote{For instance in $\eta$-deformed   $PSU(2,2|4)$ current understanding is that only scale invariance holds \cite{Arutyunov:2015mqj}.
   }?
   \item What are the corresponding affine symmetries and the exact value of the central charge at FP2?
   \item What is the spectrum of primaries for this postulated CFT at FP2? 
         \end{itemize}
These are evidently interesting challenges that we hope to return to in a future paper.  For the present we continue with our principle concern; the YB-WZ model on the integrable locus.

 \section{Poisson-Lie T-duality of the YB-WZ model} \label{s4}

Motivated by the Poisson-Lie  symmetric structure of the $\eta$-deformation, one could wonder how the YB-WZ action \eqref{eq:act} behaves under Poisson-Lie symmetry. Remarkably the YB-WZ model at the integrable locus features an example of the most simple realisation of PL. The Poisson-Lie duality transformation preserves the structure of the action, reshuffling the coupling constant in a surprisingly Busher-rule like manner. At the RG fixed point (the WZW) the action is self-dual. This section, being somewhat technical, can safely be omitted at a first reading and the reader can jump directly to the resulting ``effective'' transformation rules of the Poisson-Lie transformation of the YB-WZ model in equations \eqref{eq:plduality}.
 
When restricted to the integrable locus, the YB-WZ model admits a 1st order formulation as an ${\mathcal E}$-model \cite{Klimcik:2017ken}.  We refer the reader to the original paper for full details of this construction but note here the essential ingredients of an  ${\mathcal E}$-model, and its connection to sigma models, are:
\begin{enumerate}
\item[(i)] An even dimensional real Lie-algebra $\frak{d}$ 
\item[(ii)] An ad-invariant inner product  $( \cdot , \cdot )_\frak{d}: \frak{d}\otimes \frak{d} \to \mathbb{R}$
\item[(iii)] An idempotent involution ${\mathcal E}$ that is self-adjoint with respect to the inner product
\item[(iv)] A maximally isotropic subalgebra $\frak{h}$ (i.e.\ $(z_1, z_2)_\frak{d} = 0\;  \forall z_{1,2} \in \frak{h}$ and $ \dim \frak{h} = \frac{1}{2} \dim \frak{d} $).  
\end{enumerate}
Given the data of (i)-(iii) one can construct a 1st order action known as the ${\mathcal E}$-model. Given further (iv)   one can integrate out auxiliary fields from the  ${\mathcal E}$-model to arrive at a non-linear sigma model.  The field variables of this sigma model are sections (defined patchwise if needed) of the coset $D/H$ (with $ D,H$ the groups corresponding to $\frak{d}$ and $\frak{h}$).   If a second maximally isotropic subalgebra $\tilde{\frak{h}}$ can be found then the procedure can be repeated to yield a second non-linear sigma model on $D/\tilde{H}$ -- this is the Poisson-Lie dual.

For both the YB ($\eta$-theory) and the present case of interest, the YB-WZ theory,  the relevant algebra is $ \frak{d}= \frak{g}^\mathbb{C}$, viewed as a real Lie algebra with elements $z= x+ i y $ with $x,y \in \frak{g}$.  The addition of the WZ term requires that the inner product be modified to 
\cite{Klimcik:2017ken},
\be\label{eq:inner}
(z_1, z_2)_\frak{d} = C   \textrm{Im}~\langle e^{i\rho} z_1 , z_2 \rangle \ , 
 \ee
where the parameters used in \cite{Klimcik:2017ken} translate to, 
\be\label{eq:prmap}
C = \frac{k^2 + \Theta^2}{8\pi \Theta} \ , \quad  e^{i\rho}= -\frac{k+ i\Theta }{k- i\Theta}  = \frac{\gamma^{-}}{\gamma^{+} }  \ , 
\ee
  which are both RG invariant and match the parameters determining the (affine tower) charge algebra in the case of $SU(2)$ established in \cite{Kawaguchi:2013gma}, see also appendix \ref{a1}.   The involution $\mathcal{E}$, whose precise definition will not be illuminating for us and can be found in  eq.~(3.8) of \cite{Klimcik:2017ken},  dresses up the  swapping  of real and imaginary parts of $z\in \frak{g}^\mathbb{C}$ with parametric dependance on $\rho$ and also on $e^p = -\frac{\alpha}{\Theta}$.  So unlike the innerproduct, $\mathcal{E}$ is  RG variant.   
  
  We have two maximal isotropics given by the embeddings of $\frak{g}$:
  \be
  \begin{aligned}
  \frak{h}_\rho = \left(\R- \tan\frac{\rho}{2}( \R^2 + 1) - i  \right)\frak{g} \ ,  \\ 
  \tilde{  \frak{h}}_\rho = \left(\R-\cot\frac{\rho}{2}( \R^2 + 1) + i  \right)\frak{g}  \ . 
  \end{aligned}
  \ee
  That these are subalgebras follows immediately since  $\R$ satisfies the mCYBE and  $\R^2 + 1 $ projects  into the Cartan. That they are isotropic with respect to \eqref{eq:inner}  fixes the trigonometric functions.  Since $ \frak{h}_\rho = e^{i\rho}\frak{a}+\frak{n} $ where $\frak{a}$ and $\frak{n}$ are the corresponding algebras in the Iwasawa decomposition $D= KAN$,  we can think of $\frak{h}_\rho$ as a twisted upper triangular subalgebra and the other,    $\tilde{  \frak{h}}_\rho$, as  lower triangular. 
   Locally at least we can decompose, 
   \be
   D= H_\rho \cdot \tilde{H}_\rho    =   \tilde{H}_\rho \cdot H_\rho\, ,
   \ee    
   and because the standard Iwasawa decomposition can be modified to incorporate the twisting by $\rho$ as in     \cite{Klimcik:2017ken}  we can   identify $D \equiv G^{\mathbb{C}} =  G  \cdot \tilde{H}_\rho  = G  \cdot  H_\rho$.  Thus the cosets $D/H_\rho$ and $D/\tilde{H}_\rho$ can be identified with  $G$  and so $g\in G$ can serve as field variables on either of the two dual models.  To extract the sigma models one needs to specify projectors ${\mathcal P}$ and $\tilde{{\mathcal P}}$ such that\footnote{There is a slight simplification here of the general formulas of  \cite{Klimcik:2017ken}  since  $g \in G$ the adjoint action $ad_g$ commutes in this case with the idempotent ${\mathcal E}$.}, 
  \be 
  \textrm{Im} {\mathcal P} = \frak{h}_\rho  \ , \quad   \textrm{Ker} {\mathcal P} = (1 + {\mathcal E})\frak{d} \ , \quad     \textrm{Im} \tilde{{\mathcal P}} =\tilde{\frak{h}}_\rho  \ , \quad    \textrm{Ker} \tilde{{\mathcal P}} = (1 +  {\mathcal E})\frak{d}\, .
  \ee
 Explicitly if we let (making use of the definition of $  {\mathcal E}$ in eqs.~(3.7,3.8) of  \cite{Klimcik:2017ken}), 
 \be
 w= e^{- i \rho} + i \cosh(p)+ i e^{- i \rho }\sinh(p) = w_1 + i w_2 \ , \quad x = \frac{w_1}{w_2}  \ , 
 \ee   
  and define,
  \be
  {\mathcal O} =\R- \tan\frac{\rho}{2}(\R^2 + 1) \ , \quad   \tilde{   {\mathcal O} } =\R- \cot\frac{\rho}{2}( \R^2 + 1)    \ , 
  \ee
  then, 
  \be
  {\mathcal P}(g^{-1} dg) = \left(  {\mathcal O} - i \right) \left(   {\mathcal O} + x \right)^{-1} g^{-1} dg  \ ,  \quad  \tilde{\mathcal{P} }(g^{-1} dg) = \left(  \tilde{{\mathcal O}} + i \right) \left(  \tilde {\mathcal O} - x \right)^{-1} g^{-1} dg \,.
  \ee  
 Equipped with all of this we can now simply specify the non-linear sigma models obtained after integrating out the auxiliary fields from the ${\mathcal E}$ models.  They read, 
 \be
 \begin{aligned}
 S = S_{WZW, k}^\frak{d}[g] + \frac{k}{\pi} \int    d  \sigma d\tau    \left(  {\mathcal P} (g^{-1} \partial_+ g ) , g^{-1} \partial_- g \right)_\frak{d} \ , \\ 
\tilde{S}=  S_{WZW, k}^\frak{d}[g] + \frac{k}{\pi} \int    d  \sigma d\tau    \left(  \tilde{{\mathcal P} }(g^{-1} \partial_+ g ) , g^{-1} \partial_- g \right)_\frak{d} \ , 
\end{aligned} 
\ee   
 where we emphasise that the deformed inner product on $\frak{g}^\mathbb{C}$  of eq.~\eqref{eq:inner} is used to define the WZW models and that the term depending on the projectors has coefficient $-2$ times that of the kinetic term of the WZW model.   Using $\R^3 =- \R$ it was established in \cite{Klimcik:2017ken}  that the first of these actions matches the general model  in eq.~\eqref{eq:act} with parameters $\alpha$, $\beta$ and $\gamma$ obeying the integrable locus relation.  What of the Poisson-Lie dual theory?  After some tedious trigonometry and using the relations eq.~\eqref{eq:prmap} together with the definition of the inner product eq.~\eqref{eq:inner} one finds the action $\tilde{S}$  is also of the form of eq.~\eqref{eq:act} but the T-duality acts on the parameters as, 
 \be
 \begin{aligned}\label{eq:plduality}
 \alpha &\to  \tilde{\alpha} = \frac{k^2}{\alpha}\, ,  \\
 \beta &\to  \tilde{\beta} = - \beta\, , \\
 \gamma &\to   \tilde{\gamma} = \frac{k^2 + \alpha \gamma - \alpha^2}{\alpha} = -\frac{\beta^2}{\gamma} \, . 
  \end{aligned}
 \ee
 This is a truly elegant result; recall that $\frac{1}{k}$ plays the role of $\alpha^\prime$ so that these Poisson-Lie T-duality rules really do resemble the radial inversion of abelian T-duality.   Being canonically equivalent it must be the case that the T-dual model is also integrable, and indeed one sees that $\tilde{\alpha}, \tilde{\beta} , \tilde{\gamma}$ also sit on the integrable locus; this serves as a check of the T-duality rules.  
 
 We can see that the WZW point is rather special; it is the self-dual point of the duality transformation\footnote{Self-duality under PL of WZW models (with no deformations) was exhibited already in \cite{Klimcik:1996hp}.  }.   As remarked earlier in the RG portrait fig.~\ref{fig:RGint} there are two regions that corresponding to a real action, shaded in green and for which $\Theta^2 >0$.  The Poisson-Lie duality action simply maps the region for $\alpha<k$ one-to-one with that of $\alpha>k$; these two regions of course touch at the self-dual WZW fixed point. 
 
\noindent  The action of this T-duality on the charge algebra is also of note. It follows immediately that the RG invariant combination is transformed as,
 \be
 \Theta \to \tilde{\Theta}= \frac{k^{2}}{\Theta} \ .
 \ee
 Then we see that the quantum group parameter $q = \exp \left[ \frac{8 \pi \Theta} { k^2+\Theta^2 }\right]$ is invariant under T-duality.  However recall that the affine tower of charges (at least in the $\frak{su}(2)$ where it has been established explicitly) differs from the standard affine quantum group by a multiplicative factor between gradations of $- \frac{k+ i \Theta}{k-i \Theta}$.  This factor undergoes an S-transformation, i.e it is mapped to negative its inverse. This illustrates that whilst the T-duality rules look quite trivial, at the level of charges the canonical transformation that maps the two T-dual theories can have quite an involved action.

\section{The supersymmetric YB-WZ model} \label{s6}
This section falls a bit outside the main line of the paper but is motivated by the following observation. It is clear from the previous discussion that starting from a generic $d=2$ non-linear $\sigma $-model and requiring (classical) integrability, imposes severe restrictions on the target manifold and its metric and  torsion 3-form. However another way to restrict the allowed background geometries is by requiring the existence of extended worldsheet supersymmetries.  Indeed asking that the non-linear $ \sigma $-model exhibits $N>(1,1)$ supersymmetry introduces additional geometric structure which only exists  for particular background geometries. A hitherto unexplored terrain is the eventual relationship between integrable models on the one hand and extended supersymmetry on the other (however see \cite{Figueirido:1988ct} for some early work in this direction).

In this section we explore the possibility of having $N>(1,1)$ supersymmetry in the YB-WZ models studied in this paper. This is an interesting point in itself because if one thinks about the potential use of these integrable models as  backgrounds for type II superstrings in the NS worldsheet formulation then the existence of an $N=(2,2)$ supersymmetric extension is necessary as well. As we will see, the integrable deformations of the WZW-model studied in this paper do generically not allow for an extended supersymmetry.


Given is a non-linear sigma model with target manifold $ {\cal M}$ endowed with a metric $G$ and a closed 3-form $H$ (the torsion). Locally we write $H=dB$. Passing to an $N=(1,1)$ supersymmetric extension of the model does not require any further geometric structure. Indeed the action for the $N=(1,1)$ supersymmetric non-linear sigma model written in $N=(1,1)$ superspace is remarkably similar to the non-supersymmetric one\footnote{A brief summary of our superspace conventions can be found in appendix A.},
\begin{eqnarray}
{\cal S}=\int d^2\sigma  \,d^2\theta \, D_+X^\mu (G_{\mu \nu }+B_{\mu \nu })D_-X^\nu \,,
\end{eqnarray} 
where $X^\mu $ are some local coordinates on the target manifold.

However asking for more supersymmetry does introduce additional geometrical structure. E.g.\ $N=(2,2)$ supersymmetry requires the existence of two complex structures $ {\cal J}_+$ and $ {\cal J}_-$ which are endomorphisms of the tangent space $T\mathcal{M}$ and which are such that $(\mathcal{M},G, H, {\cal J}_\pm)$ is a bihermitian structure \cite{Gates:1984nk}, \cite{Howe:1985pm}, {\em i.e.}\ $\mathcal{M}$ is even-dimensional and the complex structures ${\cal J}_\pm$ satisfy,
\begin{enumerate}
\item $ {\cal J}_\pm ^{2} = -{\bf 1}$,
\item $\left[ X,Y \right] + {\cal J}_\pm\left[{\cal J}_\pm X,Y \right] +{\cal J}_\pm \left[ X,{\cal J}_\pm Y \right]-\left[ {\cal J}_\pm X, {\cal J}_\pm Y \right] = 0$ for all $X,Y \in T\mathcal{M}$, which is the integrability condition for the complex structures,
\item $G({\cal J}_\pm X,Y) = - G(X, {\cal J}_\pm Y)$ for all $X,Y \in T\mathcal{M}$, so $G$ is a hermitian metric with respect to both complex structures,
\item $\nabla^{(+)} {\cal J}_+ = \nabla^{(-)} {\cal J}_- = 0$ with $\nabla^{(\pm)}$ covariant derivatives which use the Bismut connections:
\begin{equation}
\Gamma^{(\pm)} = \{ \} \pm \frac{1}{2} G^{-1} H\, ,
\end{equation}
such that in a local coordinate bases,
\begin{equation}
\nabla_{\rho}{\cal J}_{\pm}^{\mu}{}_\nu = \pm\frac{1}{2}\left(G^{\kappa\lambda}H_{\lambda\rho\nu}{\cal J}_{\pm}^{\mu}{}_\kappa - G^{\mu\lambda}H_{\lambda\rho\kappa}{\cal J}_{\pm}^{\kappa}{}_\nu \right)\,,
\end{equation}
where the covariant derivative $\nabla$ in the above is taken with the Christoffel symbol as connection.
This condition is equivalent to the requirement that the exterior derivative of the two-forms $\omega_\pm (X,Y) = -G(X,{\cal J}_\pm Y)$ are given by:
\begin{equation}
\mathrm{d}\omega_{\pm}(X,Y,Z) = \mp H ({\cal J}_\pm X,{\cal J}_\pm Y, {\cal J}_\pm Z).
\end{equation}
\end{enumerate}
Using the covariant constancy of the complex structure one can rewrite the integrability condition (condition 2) as,
\begin{eqnarray}
H(X, {\cal J}_\pm Y, {\cal J}_\pm Z)+H(Y, {\cal J}_\pm Z, {\cal J}_\pm X)+H(Z, {\cal J}_\pm X, {\cal J}_\pm Y)=H(X,Y,Z)\,. \label{kop3} 
\end{eqnarray} 
Note that demanding $N=(2,0)$ or $N=(2,1)$ instead of $N=(2,2)$ supersymmetry only requires the existence of $ {\cal J}_+$ satisfying the above conditions.  
 
We now rewrite these conditions for the deformed models studied in this paper. Since at the level of the action the deformation preserves the left acting $G$ symmetry while it breaks the right acting $G$ symmetry (to its Cartan subgroup), the geometry and the $\N=(2,2)$ conditions are most naturally presented in the basis of left-invariant one-forms $u^{A}$. Given the deformed metric $G_{AB}$ eq.~\eqref{eq:metricansatz} and the torsion $H_{ABC}$ eq.~\eqref{eq:torsion}, we find that the above conditions for $\N=(2,2)$ supersymmetry translate in this basis to the following:
\begin{enumerate}
\item The first condition is simply,
\begin{eqnarray}
\J_{\pm}^{A}{}_C \J_{\pm}^{C}{}_B = -\delta^{A}_B\,.\label{fd1}
\end{eqnarray} 
\item The second condition (using the form in eq.~(\ref{kop3}))  results in:
\begin{eqnarray}
3{\cal  J}_\pm^D{}_{[A} {\cal  J}_\pm^E{}_{B}H_{C]DE}=H_{ABC}\,,\label{fd7}
\end{eqnarray} 
where $H_{ABC}$ is given in eq.~(\ref{eq:torsion}). 
\item The third condition yields, $G_{AB} = \J_{\pm}^{C}{}_A \J_{\pm}^{D}{}_B G_{CD}$ or using eq.~(\ref{eq:metricansatz}) :
\begin{equation}
\J_\pm^{C}{}_A \kappa_{CB} = -\J_\pm^{C}{}_B \kappa_{CA} - \frac{\gamma}{\alpha} \left(\J_\pm^{C}{}_B \R^{2}_{CA} + \J_\pm^{C}{}_A  \R^{2}_{CB} \right)\,.\label{fd3}
\end{equation}
\item After a little bit of work, the covariant constancy of the complex structures (the fourth condition), translates to,
\begin{equation}
u_C{}^{\mu}\partial_\mu \J_{\pm}^{A}{}_B =  \J_{\pm}^{A}{}_DM^D_\pm{}_{CB} - M^A_\pm{}_{CD}\J_{\pm}^{D}{}_B\,,\label{fd4}
\end{equation}
where,
\begin{eqnarray}
M^A_\pm{}_{BC}=\Gamma^{A}_{BC} \pm  \frac{1}{2}G^{AD}H_{DBC}\,,\label{conner}
\end{eqnarray} 
and where the spin connection $\Gamma^{A}_{BC}$ is given by eq.~\eqref{eq:spinconn}. 
Eq.~(\ref{fd4}) implies an integrability condition,
\begin{eqnarray}
{\cal J}_\pm^A{}_E \,R_\pm^E{}_{BCD}=R_\pm^A{}_{ECD}\, {\cal J}_\pm^E{}_B\,,  \label{intcur}
\end{eqnarray}  
where the curvature tensors $R_\pm$ are given by,
\begin{eqnarray}
R^{A}_\pm{}_{BCD} = M_\pm^{E}{}_{DB} M_\pm^{A}{}_{CE} -  M_\pm^{E}{}_{CB} M_\pm^{A}{}_{DE}-F_{CD}{}^{E} M_\pm^{A}{}_{EB}\,.\label{curvdef}
\end{eqnarray} 
The integrability condition eq.~(\ref{intcur}) is the requirement that the complex structures commute with the generators of the holonomy group defined by the connections in eq.~(\ref{conner}). 
\end{enumerate}

While the first three conditions are given by algebraic equations eqs.~(\ref{fd1})--(\ref{fd7}), which can be analyzed in a way similar to what was done in  \cite{Spindel:1988nh,Spindel:1988sr}, the last one, eq.~(\ref{fd4}), is involved. However, the integrabilty conditions for the latter are algebraic again and can be explicitly analyzed.

In \cite{Spindel:1988nh,Spindel:1988sr} these conditions were analyzed for the undeformed case, $\alpha =|k|$, $\beta =\gamma =0$, {\em i.e} the standard WZW model and it was found that on any even-dimensional group manifolds there exist solutions to the above equations. Let us briefly review those results. In the undeformed case the connections in eq.~(\ref{conner}) are simply,
\begin{eqnarray}
M_+^A{}_{BC}=0\,,\qquad M_-^A{}_{BC}=F_{BC}{}^A\,.
\end{eqnarray} 
With this one verifies that the curvature tensors in eq.~(\ref{curvdef}) vanish (reflecting the fact group manifolds are parallelizable), either trivially or by virtue of the Jacobi identities, and as a consequence the integrabilty conditions, eq.~(\ref{intcur}), are automatically satisfied. Turning then to eq.~(\ref{fd4}) one finds that ${\cal J}_+^A{}_B$ is constant while ${\cal J}_-^A{}_B$ satisfies,
\begin{eqnarray}
u_C{}^{\mu}\partial_\mu \J_{-}^{A}{}_B =  \J_{-}^{A}{}_D\,F_{CB}{}^D - F_{CD}{}^A\,\J_{-}^{D}{}_B\,. \label{yx11}
\end{eqnarray}  
In order to analyze the latter one introduces a group element in the adjoint representation,
\begin{eqnarray}
S^A{}_B= v_\mu ^A\,u^\mu _B\,,
\end{eqnarray} 
which can easily be shown to satisfy,
\begin{eqnarray}
u_C{}^{\mu}\partial_\mu S^A{}_B=F_{CB}{}^D\,S^A{}_D\,.
\end{eqnarray} 
Using this and eq.~(\ref{yx11}) one  shows that $S^A{}_C{\cal J}_-^C{}_D(S^{-1} )^D{}_B$ (which is ${\cal J}_-$ in the right invariant frame) is constant as well. In this way the remaining conditions for $N=(2,2)$ supersymmetry, eq.~(\ref{fd1})-(\ref{fd3}), all reduce to algebraic equations on the Lie algebra which were solved in \cite{Spindel:1988nh,Spindel:1988sr}. The result is remarkably simple: any complex structure pulled back to the Lie algebra is almost completely equivalent to a choice for a Cartan decomposition. Indeed the complex structure acts diagonally on generators corresponding to positive (negative roots) with eigenvalue $+i$ ($-i$). It maps the CSA to itself so that it squares to minus one and so that the Cartan-Killing metric restricted to the CSA is hermitian.

In the deformed case the integrability conditions eq.~(\ref{intcur}) become non-trivial and need to be  investigated first. While in principle this can be done for general groups (resulting in not particularly illuminating complex expressions) we limit ourselves in this paper to a detailed analysis of the simplest case: $SU(2)\times U(1)$.  A more systematic analysis of the relation extended supersymmetry and integrability in general is currently underway and will be reported on elsewhere. 

For $SU(2) \times U(1)$ the $\beta$ deformation is a total derivative and can be ignored in the present analysis. We choose a basis for the Lie algebra where $t_0=( \sigma _3+ i \sigma_0)/2$, $t_{\bar 0}=( \sigma _3- i \sigma_0)/2$, $t_1=( \sigma _1+ i \sigma_2)/2$ and $t_{\bar 1}=( \sigma _1- i \sigma_2)/2$. In this basis the non-vanishing components of the Cartan-Killing metric are given by $\kappa_{0\bar 0}= \kappa_{1\bar 1}=1$ and those of $ {\cal R}$ by $ {\cal R}^1{}_1=- {\cal R}^{\bar 1}{}_{\bar 1}=i$. The non-vanishing components of the deformed metric in the left invariant frame, eq.~(\ref{eq:metricansatz}), are $G_{0\bar 0}= \alpha $ and $G_{1\bar 1}= \alpha -\gamma $. For the torsion, eq.~(\ref{eq:torsion}), we get $H_{01\bar 1}=H_{\bar 0 1\bar 1}=i\,k$. The hermiticity condition eq.~(\ref{fd3}) and the integrability condition eq.~(\ref{intcur}) are both linear in the complex structures and as a consequence are easily analyzed. The hermiticity condition eliminates 10 of the 16 components of each complex structure. A straightforward but somewhat tedious calculation shows the following result for the integrability condition:
\begin{enumerate}
\item It is identically satisfied without any further conditions if $ \alpha =|k|$ and $\gamma =0$. This is just the undeformed $SU(2)\times U(1)$ WZW model known to be $N=(2,2)$ supersymmetric (in fact it is even $N=(4,4)$ supersymmetric ). 
\item It is satisfied if $\alpha =|k|$ and only ${\cal J}_\pm^0{}_0= - {\cal J}_\pm^{\bar 0}{}_{\bar 0}$ and  ${\cal J}_\pm^1{}_1= - {\cal J}_\pm^{\bar 1}{}_{\bar 1}$ are non-vanishing.
\item Otherwise, for generic values of $\alpha $, $ \gamma $ and $k$ it has {\em no} solutions. 
\end{enumerate}     
So we can conclude that in general the deformed $SU(2)\times U(1)$ YB-WZ model does not allow for an $N=(2,2)$ supersymmetric extension. Remains of course case 2 in the above. 

From now on we take $\alpha =|k|$. Checking eq.~(\ref{yx11}) one finds that only a vanishing $ {\cal J}_-$ is consistent with eqs.~(\ref{fd1}) and (\ref{fd7}) while $ {\cal J}_+$ is constant and its non-vanishing components are given by e.g.\ ${\cal J}_+^{0}{}_{0}=-{\cal J}_+^{\bar 0}{}_{\bar 0}={\cal J}_+^{1}{}_{1}=-{\cal J}_+^{\bar 1}{}_{\bar 1}=i$.  This choice for ${\cal J}_+$ also satisfies eqs.~(\ref{fd1}) and (\ref{fd7}). So we conclude that the model is indeed $N=(2,1)$ or $N=(2,0)$ supersymmetric but does not allow for $N=(2,2)$ supersymmetry. 

To end this section we formulate this model in $N=(2,1)$ superspace thereby making the $N=(2,1)$ supersymmetry explicit. In general one starts with a set of $N=(2,1)$ superfields $z^\alpha $ and $z^{\bar \alpha }$ satisfying the constraints $\hat D_+ z^ \alpha =+i\,D_+ z^ \alpha $ and $ \hat D_+ z^ {\bar\alpha} =-i\,D_+ z^ {\bar\alpha} $ which are a consequence of the fact that the non-vanishing components of the complex structure ${\cal J}_+$ are ${\cal J}_+^\alpha {}_\beta =+i \,\delta ^\alpha _ \beta $ and ${\cal J}_+^{\bar \alpha} {}_{\bar\beta} =-i \,\delta ^{\bar \alpha }_ {\bar\beta} $. The action is expressed in terms of a vector on the target manifold $(i\,V_\alpha,-i\, V_{\bar \alpha })$,
\begin{eqnarray}
{\cal S}=\frac 1 2 \,\int d^2 \sigma\, d^2 \theta\, d \hat\theta  {}^+ \Big(
i\,V_\alpha \,D_-z^\alpha -i\,V_{\bar \alpha }\,D_-z^{\bar \alpha }
\Big)\,.\label{n21action}
\end{eqnarray}  
Passing to $N=(1,1)$ superspace,
\begin{eqnarray}
{\cal S}&=&\int d^2 \sigma\, d^2 \theta\Big(
G_{\alpha \bar \beta }(D_+z^ \alpha D_-z^{\bar \beta }+D_+z^ {\bar \beta } D_-z^{\alpha })+ \nonumber\\
&&\qquad
B_{\alpha \bar \beta }(D_+z^ \alpha D_-z^{\bar \beta }-D_+z^ {\bar \beta } D_-z^{\alpha })
\Big)\,,
\end{eqnarray} 
one identifies the metric,
\begin{eqnarray}
G_{ \alpha \bar \beta }=G_{  \bar \beta \alpha }=\frac 1 2\,\Big(\partial _ \alpha V_{\bar \beta }+ \partial _ {\bar \beta } V_{ \alpha  }\Big)\,,
\label{ggg2}
\end{eqnarray} 
and the Kalb-Ramond 2-form,
\begin{eqnarray}
B_{ \alpha \bar \beta }=-B_{  \bar \beta \alpha }=\frac 1 2 \Big(\partial _ \alpha V_{\bar \beta }- \partial _ {\bar \beta } V_{ \alpha  }\Big)\,. 
\label{bbb2}
\end{eqnarray} 

Now let us apply this to the deformed $SU(2)\times U(1)$ model where $ \alpha = |k|$. The group element $g\in SU(2)\times U(1)$ is parameterized in a standard way by,
\begin{eqnarray}
g=e^{\frac i 2\rho }\,\left(
\begin{array}{cc} e^{\frac i 2(\varphi_1+\varphi_2)}\, \cos\frac \psi 2 & e^{\frac i 2 (\varphi_1-\varphi _2)}\,\sin \frac \psi 2 \\
-e^{-\frac i 2 (\varphi_1-\varphi _2)}\,\sin\frac \psi 2 &e^{-\frac i 2 (\varphi_1+\varphi _2)}\, \cos\frac \psi 2\end{array}\right)\,,
\end{eqnarray} 	
where,
\begin{eqnarray}
0\leq \psi\leq \pi\,,\qquad \varphi_1\in\IR\mbox{ mod } 2\pi\,,\qquad \varphi_2\,,\,\rho\in\IR\mbox{ mod } 4\pi\,.
\end{eqnarray} 
We introduce complex coordinates $z^{\alpha} = (z,w)$ and $\bar{z}^{\bar{\alpha}}=(\bar{z}, \bar{w})$ with $\alpha, \bar{\alpha}=\{1,2\}$ such that $ {\cal J}_+$ acts as $+i$ on $dz$ and $ dw$. The complex structure in this case is exactly the same as the one studied originally in \cite{Rocek:1991vk} (for a more detailed treatment see \cite{Sevrin:2011mc}), so we can use the results obtained there to write the group element in terms of the complex coordinates as,
\begin{eqnarray}
g=(z\bar z+w\bar w)^{-\frac 1 2(1+i)}\left(
\begin{array}{cc} \bar w & z \\
-\bar z &w\end{array}\right)\,,
\end{eqnarray} 
where the complex coordinates are related to the original coordinates as,
\begin{eqnarray}
z=e^{-\frac 1 2 \rho}\,e^{\frac i 2(\varphi_1-\varphi_2)}\,\sin \frac \psi 2 \,,\qquad
w=e^{-\frac 1 2 \rho}\,e^{-\frac i 2(\varphi_1+\varphi_2)}\,\cos \frac \psi 2 \,.
\end{eqnarray} 
Note that in the undeformed case, which allows for $N=(2,2)$ supersymmetry, $z$ and $w$ are the chiral and the twisted chiral $N=(2,2)$ superfield resp.\   \cite{Rocek:1991vk}. In the undeformed case we can readily derive the $N=(2,1)$ vector $V_z$ and $V_w$ appearing in the action eq.~(\ref{n21action}) as it directly descends from the generalized $N=(2,2)$ K\"ahler potential $K$ obtained in \cite{Rocek:1991vk}, 
\begin{eqnarray}
K(z,\bar z, w, \bar w)=-k\,\Big(\frac 1 2 \,(\ln w\bar w)^2+Li_2\left(-\frac{z\bar z}{w\bar w}\right)\Big)\,,
\end{eqnarray} 
from which we get,
\begin{eqnarray}
&&V^0_z = +\partial _z K=\frac{k}{z}\, \ln\Big(1 + \frac{z \bar z}{w \bar w}  \Big)\,, \nonumber\\
&&V^0_w = -\partial _w K= \frac{k}{w}\,\ln( z \bar {z } + w \bar{w} )\,,\label{vvv0}
\end{eqnarray} 
where the upper index $0$ on $V$ points to the fact that we are dealing with the undeformed case $\alpha =|k|$ and $\gamma =0$.

In order to extend this to the deformed case, i.e.\ $\gamma \neq 0$, we first rewrite the deformed geometry in terms of complex coordinates,
\begin{eqnarray}
&&ds^2 = \frac{k}{z \bar{z} + w \bar{w} }\, \Big( d z\, d \bar {z}+d w \,d\bar{w}  \Big) - \frac{\gamma}{(z \bar{z} + w \bar{w})^2}\, \Big(w\,dz -z\, dw \Big)\Big(\bar w\, d \bar z-\bar z \,d\bar w \Big)\,, \nonumber\\
&&H= \frac{k}{ (z \bar{z} + w \bar{w} )^2}\,  \Big( d z \wedge d \bar z  \wedge (\bar w \,d w  -w \,d \bar w) +d w \wedge d \bar w \wedge (\bar z\, d z  -z\,  d \bar z)   \Big)\,,\label{goe6}
\end{eqnarray} 
where we put $\alpha =k$.
From the expression for the torsion one gets the Kalb-Ramond 2-form as well,
\begin{eqnarray}
B= \frac{k}{z\bar z +w\bar w}\,\Big( \frac{\bar z}{\bar w}\,d z \wedge d\bar w-\frac{ z}{ w}\,d w \wedge d\bar z
\Big)\,.\label{goe7}
\end{eqnarray} 
From this we  obtain $V_z$ and $V_w$,
\begin{eqnarray}
V_z =V^0_z  + \frac{\gamma}{z } \,\frac{w \bar{w} } {z \bar{z}+ w \bar{w}}    \, ,  \qquad V_w = V^0_w + \frac{\gamma}{w } \,\frac{z \bar{z} } { z \bar{z}+w \bar{w}}    \,  ,\label{vvv5}
\end{eqnarray} 
where $V^0_z$ and $V^0_w$ were given in eq.~(\ref{vvv0}). Using eqs.~(\ref{ggg2})  and (\ref{bbb2}) one verifies that eq.~(\ref{vvv5}) indeed reproduces eq.~(\ref{goe6})  and (\ref{goe7}). Combining eq.~(\ref{vvv5}) with eq.~(\ref{n21action}) gives the action of the deformed theory explicitly in $N=(2,1)$ superspace.

Concluding: as the  generic $SU(2)\times U(1)$ YB-WZ model does not allow for $N=(2,2)$ supersymmetry, it looks highly improbable that deformed models for other groups would allow for $N=(2,2)$ supersymmetry. Even when only requiring $N=(2,0)$ or $N=(2,1)$ supersymmetry, one finds that this is only possible for specific values of the deformation parameters. However, it is important to note that the above derivation is based on the canonical form of the $\R$-matrix. There is still an $\mathrm{GL}(2,\mathbb{C})$ freedom on the CSA directions of $\R$ which,  together with the possibility of going beyond a single Yang-Baxter to a bi-Yang-Baxter deformation, could still reveal extended supersymmetry (but since these geometries are more complicated it would seem unlikely that they are more amenable to supersymmetries).

\section{Summary, conclusions and outlook} \label{s7}

In this paper we investigated various properties of the Yang-Baxter deformation of the Principal Chiral model with a Wess-Zumino term introduced in \cite{Delduc:2014uaa}.

As the undeformed model, the WZW model, exhibits rather unique features at the quantum level, we made a one-loop renormalisation group analysis of this class of models. For general groups and for generic values of the deformation parameters, the RG flow drives the theory outside the classical sigma model ansatz given in eq.~(\ref{eq:metricansatz}) and (\ref{eq:torsion}).  However, when the classical integrability condition is invoked, the renormalisation does remain within the sigma model ansatz and moreover the integrability condition is preserved along the RG flow.  The fact that a very quantum property--the RG equations--are sensitive to the consideration of classical integrability is rather suggestive.  It is therefore natural to conjecture that these models are quantum mechanically integrable. However, the non-ultralocal property of such theories precludes a direct application of the Quantum Inverse Scattering method. It would be very interesting to examine how the alleviation approach, used in the context of the related $\lambda$-models \cite{Appadu:2017fff},  might be applied here in order to unravel the quantum S-matrix.  
 
 Another interesting aspect is that the WZW model is the IR fixed point; in comparison the integrable $\lambda$-deformed WZW has the CFT situated as an UV fixed point.   This model then seems closer in spirit to the irrelevant double trace integrable deformations of 2d CFTs constructed recently in  \cite{Smirnov:2016lqw}.  Recently $\lambda$ type deformations have been studied in the context of $G_{k}\times G_{l}/ G_{k+l}$ coset theories \cite{Sfetsos:2017sep}; curiously there the CFT is recovered as an IR fixed point in the same way as we have here.

 An unanticipated feature of this class of models is that when restricting to simply laced groups but staying outside of the integrable locus, we found a second fixed point of the one-loop $\beta$-functions which is UV with respect to the IR WZW model.  Around this fixed point, the curvatures of the target space geometry are small leading us to anticipate that the existence of this fixed point is robust to higher loops.  However,  at this fixed point a number of the currents have wrong sign kinetic terms.  A conservative view would be to discard this as non-physical but this then begs the question of the UV completion of the deformation we are considering.  Tentatively we might suppose that the fixed point corresponds to a non-unitary CFT and that we have an exotic RG flow from this in the UV to the WZW in the IR.  Comparable flows have been discovered in the context of minimal models.   Needless to say it would be interesting to examine this more robustly.  A technique that might help here could be to rephrase the entire discussion of these theories in terms of the free field representations of WZW models.

An obvious exercise which remains to be done is an RG analysis of the integrable models introduced in \cite{Delduc:2017fib} that incorporate both bi-Yang-Baxter deformations and TST transformations. We expect this to be significantly more involved than the analysis performed in the current paper as the deformations in \cite{Delduc:2017fib} destroy both the left and right acting group symmetry rendering the choice of a good basis to calculate the $\beta $-functions non-trivial. 

An appealing feature of the landscape of $\eta$, Yang-Baxter and $\lambda$ deformations is that they provide tractable examples of sigma models that are Poisson-Lie T-dualisable.  The theories considered here also share this feature; in fact the Poisson-Lie duality (which normally results in quite convoluted geometries) has a remarkably simple form.  It results in a set of ``Buscher rules'' that resemble Abelian T-duality in that coupling constants are simply inverted.  We see quite explicitly the compability of Poisson-Lie duality and RG flow and in particular   we find that the self-dual point of the duality and the fixed point of RG are coincident. At this self-dual point the symmetries are enhanced and the theory becomes the WZW CFT.   With the understanding that the Heisenberg anti-ferromagnetic $XXX_{\frac{k}{2}}$ chain has a gapless regime in the same universality class as the $SU(2)_k$ WZW model \cite{Affleck:1987ch} an intriguing question is whether this PL duality  can also be given an interpretation in spin-chains.

Finally we studied the possibility of supersymmetrising these models. As for any non-linear sigma model in two dimensions an $N=(1,1)$ supersymmetric extension is always possible. Going beyond $N=(1,1)$ requires extra geometric structure, in particular every additional supersymmetry requires the existence of a complex structure satisfying various properties outlined in section \ref{s6} of this paper. Compared to the undeformed WZW model these conditions turn out to be rather involved. We solved them explictely in the simplest non-trivial example: $SU(2)\times U(1)$. For generic values of the deformation parameters no supersymmetry beyond $N=(1,1)$ is allowed. For the particular case where the deformation parameter $\alpha $, defined in eq.~(\ref{eq:act}), satisfies $\alpha =|k|$ with $k$ the level of the WZ term an $N=(2,1)$ extension is still possible while $N=(2,2)$ is forbidden. We provided the manifest supersymmetric formulation of this model in $N=(2,1)$ superspace. 

The above analysis showed no obvious connection between integrability and the existence of extended supersymmetries (perhaps this is not so surprising, see e.g.\ \cite{Figueirido:1988ct}). A useful exercise in this context would be the following. All bi-hermitian complex surfaces have been classified \cite{aposto}. Those with the topology of $SU(2)\times U(1)$ are the primary Hopf surfaces. A detailed analysis of the $N=(2,2)$ superspace formulations of those models combined with their integrability properties would be most interesting, in particular a characterization of the notion of integrability directly in $N=(2,2)$ superspace would be quite exciting. In view of the results obtained in the current paper we expect that if a connection between extended supersymmetry and integrability can be obtained it would probably not fall in the class of the models introduced in \cite{Delduc:2014uaa}, however other possibilities remain, e.g.\ the models developed in \cite{Delduc:2017fib} and through the inclusion of an action on the Cartan in the ${\cal R}$-matrix. We will come back to this issue in a future publication. 

 \section*{Acknowledgments}

\noindent
DCT is supported by a Royal Society University Research Fellowship {\em Generalised Dualities in String Theory and Holography} URF 150185 and in part by STFC grant ST/P00055X/1. This work is supported in part by the Belgian Federal Science Policy Office through the Interuniversity Attraction Pole P7/37, and in part by the ``FWO-Vlaanderen'' through the project G020714N and two ``aspirant'' fellowships (SD and SD), and by the Vrije Universiteit Brussel through the Strategic Research Program ``High-Energy Physics''.  SD and SD would like to thank Swansea University for hospitality during a visit in which part of this research was conducted.   We thank Vestislav Apostolov, Benjamin Doyon, Tim Hollowood, Chris Hull, Ctirad Klimcik, Prem Kumar, Martin Ro\v cek and Kostas Sfetsos for useful conversations/communications that aided this project. 
 \appendix
\appendixpage
 
\section{Conventions}\label{a0}
Let us  establish our conventions. In this article we consider only semi-simple Lie groups $G$.  For the corresponding Lie algebra $\frak{g}$ we pick a basis of Hermitian generators,
 \be
 [T_A ,T_B] = i \,F_{AB}{}^C T_C  \ ,
 \ee
 where $F_{AB}{}^C$ are the structure constants which satisfy the Jacobi identity:
  \begin{equation}
 F_{AB}{}^D F_{DC}{}^E + F_{CA}{}^D F_{DB}{}^E + F_{BC}{}^D F_{DA}{}^E = 0\,.
 \end{equation}
 We denote by  $\langle \cdot , \cdot  \rangle : \frak{g}\times \frak{g} \rightarrow \mathbb{R}$   the ad-invariant Cartan-Killing form on   $\frak{g}$ whose components are  $  \langle T_A ,T_B \rangle = \frac{1}{x_R} Tr(T_A T_B)=   \kappa_{AB}$ (with $x_R$ the index of the representation $R$).  In particular one gets for the adjoint representation,
 \begin{eqnarray}
F_{AC}{}^DF_{BD}{}^C=-c_G\, \kappa_{AB}\,,\label{CKmet}
\end{eqnarray} 
 with $c_G=2 h^\vee$ where $h^\vee$ is the dual Coxeter number of the group.
 
Going now to a Cartan-Weyl basis where we call the generators in the Cartan subalgebra (CSA) $H_m$, the generators corresponding to positive (negative) roots $T_a$ ($T_{\bar a}$), where we have $[H_m,T_a]= a_m\,T_a$ and $[H_m,T_{\bar a}]= -a_m\,T_{\bar a}$.  Using this one immediately gets from eq.~(\ref{CKmet}),
\begin{eqnarray}
\kappa_{mn}= \frac{1}{h^\vee}\sum_a a_ma_n\,,
\end{eqnarray}  
where the sum runs over the positive roots. With this we define the length squared of a root $\vec a$ by\footnote{$\kappa^{mn}$ is the inverse of $\kappa_{mn}$.} $\vec a\cdot \vec a=a_m\kappa^{mn}a_n$.  With our choice for the normalization of the Cartan-Killing form the length squared of the long roots is always 2 and for the non-simply laced groups the length squared of the short roots is either 1 or 1/3.
 
 We define left-invariant forms $u = -i u^A  T_A = g^{-1} d g$   which thus obey $du^A = -\frac{1}{2} F_{BC}{}^A u^B \wedge u^C$ whilst  right-invariant forms  $v = -i v^A  T_A =  d gg^{-1} $  obey  $dv^A = + \frac{1}{2} F_{BC}{}^A v^B \wedge v^C$. 
 The Wess-Zumino-Witten  action \cite{Witten:1983ar} is, 
   \be
  S = -\frac{k}{2\pi}\int_\Sigma  d  \sigma d\tau    \langle g^{-1} \partial_+ g , g^{-1} \partial_- g \rangle + \frac{  k}{24\pi }       \int_{M_3}     \langle   \bar g^{-1} d\bar g, [\bar g^{-1} d\bar g,\bar g^{-1} d\bar g]  \rangle \ , 
   \ee
 in which  $g:\Sigma \rightarrow G$ and with $\bar{g}$ the extension of $g$ into $M_3$ such that $\partial M_3 = \Sigma$.  We adopt light-cone coordinates      $\sigma^\pm = \tau \pm \sigma $.   For compact $G$, and demanding that the action is insensitive to the choice of action, requires $k\in \mathbb{Z}$. 
 
 In section \ref{s6} we deal with non-linear sigma models in $N=(1,1)$ and $N=(2,1)$ superspace. Let us briefly review some of the notations appearing there and refer to e.g.\ \cite{Sevrin:2011mc} for more details. Denoting for this section the bosonic worldsheet light-cone coordinates by,
\begin{eqnarray}
\sigma ^\pp= \tau + \sigma\, ,\qquad \sigma ^== \tau - \sigma \,,
\end{eqnarray}
and the $N=(1,1)$ (real) fermionic coordinates by $ \theta ^+$ and $ \theta ^-$, we introduce the
fermionic derivatives which satisfy,
\begin{eqnarray}
D_+^2= - \frac{i}{2}\, \partial _\pp \,,\qquad D_-^2=- \frac{i}{2}\, \partial _= \,,
\qquad \{D_+,D_-\}=0\,.\label{App2}
\end{eqnarray}
The $N=(1,1)$ integration measure is given by,
\begin{eqnarray}
\int d^ 2 \sigma \,d^2 \theta =\int d\tau \,d \sigma \,D_+D_-\,.
\end{eqnarray}
Passing from $N=(1,1)$ to $ N=(2,1)$ superspace requires
the introduction of one more real fermionic coordinates $ \hat \theta ^+$ where the corresponding fermionic derivative satisfies,
\begin{eqnarray}
\hat D_+^2= - \frac{i}{2} \,\partial _\pp \,,
\end{eqnarray}
and all other -- except for (\ref{App2}) -- (anti-)commutators do vanish.
The $N=(2,1)$ integration measure is,
\begin{eqnarray}
\int d^2 \sigma \,d^2 \theta \, d \hat \theta^+ =
\int d \tau\, d \sigma \,D_+D_-\, \hat D_+ \,.
\end{eqnarray}
 
 \section{Charges in $SU(2)$}\label{a1}
  
 In this appendix we review the construction \cite{Kawaguchi:2013gma}  of charges satisfying a quantum group algebra for the case of $\frak{g}= \frak{su}(2)$ paying rather careful attention to the normalisation of canonical momenta so as to obtain the quantum group parameters expressed in terms of RG invariant quantities.

In this appendix we use $\frak{su}(2)$ generators   $[T^\pm , T^3] = i T^\pm$, $[T^+ , T^- ]= - i T^3$ and  define components of the left invariant one-forms via $g^{-1} dg \equiv u_+ T^+ + u_- T^- + u_3 T^3$.

To orientate ourselves we begin with the Lagrangian  eq.~\eqref{eq:act}  specialised to the case of the $\eta$-deformation, i.e.\ $ \alpha = \frac{1}{\tau} \ , \ \beta = \frac{\eta}{1+\eta^2} \frac{1}{\tau}  \ , \ \gamma =   \frac{\eta^2}{1+\eta^2} \frac{1}{\tau}$ with $k=0$ incorporating some of the key points of \cite{Kawaguchi:2012ve,Delduc:2013fga}. Let us define some at first sight non-obvious currents,
\be
j_\pm = -\frac{1}{2} \frac{\eta}{1+\eta^2} \frac{1}{\Sigma} \left(\eta\, u_{\sigma\pm } \pm i u_{\tau \pm }  \right) \,, \quad j_3= \frac{\eta}{2\Sigma} u_{\tau 3}   \, , 
\ee
in which $\Sigma = 4 \pi \tau \eta$.  These have simple Poisson brackets,
\be
\begin{aligned}
\{ j_3 (\sigma_1) , j_3(\sigma_2) \} &=  0  \ , \\
\{ j_\pm (\sigma_1) , j_3(\sigma_2) \} &= \pm i j_\pm (\sigma_2) \delta(\sigma_1- \sigma_2) \ ,  \\ 
\{ j_\pm (\sigma_1) , j_\mp(\sigma_2) \} &= \mp i  j_3(\sigma_2) \delta(\sigma_1- \sigma_2)  \ . 
 \end{aligned}
\ee
That these are indeed the correct objects to work with becomes evident if we look at the Lax connection of eq.~\eqref{eq:lax}.  Recall that the path-ordered exponential integral of the spatial component of the Lax defines conserved charges.  Expanding around particular values of the spectral parameter gives expressions for the charges.  In particular if we expand the gauge transformed Lax $\Lax^g(z) = g^{-1}  \Lax_\sigma(z) g - g^{-1} \partial_\sigma g$ around certain points $z= \pm i \eta$ --these correspond to poles in the twist function of the Maillet r/s kernels-- we find  that these currents occur naturally as, 
\be
\Lax^g(z=\mp i \eta)   = 4\Sigma j_\pm T^\pm \pm 2 i \Sigma j_3 T^3 \ .
\ee
Using the fact that the Cartan element can be factored in the path ordered exponential occurring in the monodromy matrix \cite{Kawaguchi:2012gp} one is led to construct  (non-local) currents,
\be
\begin{aligned}
\frak{J}_+(\sigma ,\tau) &= j_+(\sigma ,\tau) \exp\left[- 2 \Sigma \int_{\sigma}^\infty  j_3 (\hat{\sigma}  ,\tau)  d\hat{\sigma}  \right] \ , \\  
\frak{J}_-(\sigma ,\tau) &= j_-(\sigma ,\tau) \exp\left[ 2 \Sigma \int_{-\infty}^\sigma  j_3 (\hat{\sigma}  ,\tau)  d\hat{\sigma}  \right] \ , \\  
\frak{J}_3(\sigma ,\tau) &= j_3(\sigma ,\tau) \ . 
 \end{aligned}
\ee 
The equations of motion imply $\partial_\tau \frak{J} = \partial_\sigma \tilde{ \frak{J}}$ for some $\tilde{ \frak{J}}$ whose explicit form is not important to us and thus that the charges $\frak{Q} = \int_{-\infty}^\infty \frak{J} d\sigma$   are conserved subject to standard boundary fall off.  The Poisson brackets give,
\be\begin{aligned}
\{ \frak{J}_+(\sigma_1), \frak{J}_-(\sigma_2) \} 
&=& \frac{i}{4   \Sigma} \, \delta(\sigma_1- \sigma_2) \,    \partial_{\sigma_2}    \exp\left[ - 2 \Sigma \left(\int_{\sigma_2}^\infty  -  \int_{-\infty}^{\sigma_2}\right)  j_3 (\hat{\sigma}) d\hat{\sigma}  \right]  \ ,
 \end{aligned} 
\ee
 where we note that ``cross terms'' involving the non-local exponentials cancel.   Thus one finds that, with suitable normalisation, 
 \be\label{eq:quantumgroup}
 \{ \frak{Q}_+, \frak{Q}_- \}  =   i\, \frac{q^{\frak{Q}_3} - q^{- \frak{Q}_3}  }{q -q^{-1} } \ , \quad  \{ \frak{Q}_\pm, \frak{Q}_3 \}  =  \pm i \,\frak{Q}_\pm \ , \quad  
 q =  e^{2 \Sigma} \ .
 \ee
  
  Now we turn to the full theory including the WZ  term.  For this case we have the definitions, 
  \be
  j_\pm = - \frac{k\mp i \Theta}{8 \pi ( \alpha^2 + \Theta^2 ) } \left(( \pm i \alpha^2 + k \Theta ) u_{\sigma \pm} + \alpha(\pm i k +\Theta )u_{\tau \pm}  \right)  \ , \quad j_3 = \frac{1}{8\pi} \left(k  u_{ \sigma 3} + \alpha   u_{ \tau 3}  \right) \ ,
  \ee
  which obey a non-ultralocal algebra, 
  \be
\begin{aligned}
\{ j_3 (\sigma_1) , j_3(\sigma_2) \} &=   - \frac{k}{4\pi} \partial_{\sigma_1} \delta(\sigma_1- \sigma_2)  \ , \\
\{ j_\pm (\sigma_1) , j_3(\sigma_2) \} &= \pm i j_\pm (\sigma_2) \delta(\sigma_1- \sigma_2) \ ,  \\ 
\{ j_\pm (\sigma_1) , j_\mp(\sigma_2) \} &= \mp i  j_3(\sigma_2) \delta(\sigma_1- \sigma_2)- \frac{k}{4\pi} \partial_{\sigma_1} \delta(\sigma_1- \sigma_2)  \ , 
 \end{aligned}
\ee
and from which we can build in the same way as above {\em mutatis mutandis}  (non-local) conserved currents as,
  \be
\begin{aligned}
\frak{J}_+(\sigma ,\tau) &= j_+(\sigma ,\tau) \exp\left[ \frac{8 \pi}{\Theta - i k } \int_{\sigma}^\infty  j_3 (\hat{\sigma}  ,\tau)  d\hat{\sigma}  \right] \ , \\  
\frak{J}_-(\sigma ,\tau) &= j_-(\sigma ,\tau) \exp\left[  -\frac{8 \pi}{\Theta + i k }  \int_{-\infty}^\sigma  j_3 (\hat{\sigma}  ,\tau)  d\hat{\sigma}  \right] \ , \\  
\frak{J}_3(\sigma ,\tau) &= j_3(\sigma ,\tau) \ . 
 \end{aligned}
\ee 
At the WZW fixed point ($\alpha =|k| , \Theta= 0$) these currents just reduce to the currents generating the right acting affine $\widehat{\frak{su}(2)}$.  As with the case above these currents appear in the gauge transformed Lax expanded around the poles of its twist function, i.e., 
\be
\Lax^g\left(z= \frac{\mp i k \Theta + \alpha^2 }{ k \alpha +\mp i \Theta \alpha }   \right)   =  \frac{16\pi  \Theta }{k^2 + \Theta^2}   j_\pm T^\pm    + \frac{8 \pi}{k\mp i \Theta } j_3 T^3 \ .
\ee
 For completeness we make the identification with the parameters $\gamma_+ $ and $\gamma_-$ used in the analysis of    \cite{Kawaguchi:2013gma}:
  \be
 \gamma_+ = \frac{ 8i  \pi}{ k - i \Theta  }  \ , \quad  \gamma_- = - \frac{ 8 \pi i }{ k + i \Theta  } \,,
  \ee
  such that, 
   \be
\frac{ \gamma_- }{ \gamma_+ } =- \frac{ k - i \Theta   }{k + i  \Theta   }  \ , \quad  \Delta = \frac{\gamma^+ + \gamma^- }{2} = \frac{- 8 \pi \Theta}{\Theta^2 + k^2}\,.
  \ee
Even though the currents have a non-ultra-local algebra, the charge algebra is not ambiguous   \cite{Kawaguchi:2013gma}  (there is no order of limits problem in regulating the spatial integrals) and the commutator of charges (up to overall normalisations of $\frak{Q}^\pm$) still obeys eq.~\eqref{eq:quantumgroup}   with $q= e^{-\Delta}$. 

\section{Properties of ${\cal R}$ }\label{s:properties}
We collate here a number of identities used in the massaging of the calculation of the $\beta$-functions. The strategy of deriving these identities is practically always the same: we expand the mCYBE eq.~\eqref{eq:mcybe} or related versions in the generators $T_A$ of the Lie algebra and contract two free indices with two from the structure constants $F_{AB}{}^{C}$ or from $F_{AB}{}^D \R^{C}{}_{D}$.

For completeness, we repeat here the mCYBE:
 \be 
[{\cal R} x,  {\cal R} y]  - {\cal R} \left( [x, {\cal R} y ] + [ {\cal R}x, y] \right)  = [x, y] \quad  \forall x, y \in \frak{g}\,  .
 \ee 
 From this we can derive a related identity,
\begin{equation}
\left[ \R^{2}x,\R y\right] - \left[ \R x, \R^{2}y \right] = \R \left( \left[\R^{2}x, y \right] - \left[x,\R^{2}y \right]\right) + \left[\R x,y \right] - \left[ x, \R y\right],
\end{equation}
 and using $\R^{3}=-\R$ we can also derive:
\begin{align}\label{eq:ybr2}
&\left[\R^{2} x, \R^{2}y \right] = \R^{2}\left(\left[\R^{2}x,y \right] +\left[x, \R^{2}y \right]  \right) + (1+2\R^{2})\left[x,y\right],\\
&\left[ \R^{2}x,\R y\right] + \left[ \R x, \R^{2}y \right] =  \R \left(\left[\R^{2}x, y \right] + \left[x,\R^{2}y \right] +2\left[x,y \right]\right)+ \R^{2}\left(\left[\R x,y \right] + \left[ x, \R y\right]\right), 
\end{align}
for all $ x, y \in \frak{g}  $. This gives the following (non-exhaustive) list of properties of the  $\R$-matrix all of which were used in the derivation of the $\beta$-functions:
\begingroup
\allowdisplaybreaks
\begin{align}
&\R^{D}{}_{A}\R^{E}{}_{B} F_{DEC} +\R^{D}{}_{B}\R^{E}{}_{C} F_{DEA} +\R^{D}{}_{C}\R^{E}{}_{A} F_{DEB} -F_{ABC}= 0\,,\\
& \RR^{D}{}_{A} \R^{E}{}_{B} F_{DE}{}^C + \RR^{D}{}_{A}\R^{C}{}_{E} F_{BD}{}^E + \R^{D}{}_{A} F_{BD}{}^C + (A\leftrightarrow B) = 0\,,  \\
&\begin{multlined}[b][.9\textwidth] \RR^{D}{}_{A}\RR^{E}{}_{B} F_{DEC} - \RR^{D}{}_{C} \RR^{E}{}_{A} F_{DEB}-\RR^{D}{}_{B} \RR^{E}{}_{C} F_{DEA} \\- 2\RR^{E}{}_{C} F_{ABE} -F_{ABC}=0\,,\end{multlined} \\
& \R^{C}{}_{E} \R^{F}{}_{D} F_{CA}{}^{D} F_{FB}{}^{E} + 2 \R^{C}{}_{E} \R^{F}{}_{B} F_{CA}{}^{D} F_{DF}{}^{E} - c_G \kappa_{AB}=0\,,\\
& \RR^{D}{}_{A}\R^{E}{}_{F} F_{DE}{}^{C}F_{CB}{}^{F} = \R^{C}{}_{D} F_{AE}{}^{D}F_{CB}{}^{E},\\
& \RR^{E}{}_{C} \R^{F}{}_{A} F_{EF}{}^{D} F_{DB}{}^C + \RR^{E}{}_{C} \R^{F}{}_{D} F_{AE}{}^D F_{FB}{}^C + \R^{C}{}_{D} F_{AE}{}^D F_{CB}{}^E +c_G \R_{AB} = 0\,,\\
& \RR^{D}{}_{F} \R^{C}{}_{E} F_{CB}{}^{F} F_{AD}{}^{E} -  \RR^{D}{}_{F} \R^{C}{}_{A} F_{EB}{}^{F} F_{CD}{}^{E} + \R^{C}{}_{D} F_{AE}{}^{D} F_{CB}{}^{E} + c_G \R_{AB} = 0\,,\\
&\RR^{D}{}_{F} \RR^{E}{}_{C} F_{DA}{}^{C} F_{BE}{}^{F} + 2 \RR^{E}{}_{C} F_{DA}{}^{C}F_{BE}{}^{D} +c_G \kappa_{AB} = 0\,,\\
&\RR^{D}{}_{F} \RR^{E}{}_{C} F_{EA}{}^{F} F_{BD}{}^{C} +2 \RR^{E}{}_{C} \RR^{D}{}_{A} F_{DE}{}^{F}  F_{BF}{}^{C} - 2 c_G \RR_{AB} - c_G \kappa_{AB} = 0\,,\\
&\begin{multlined}[b][.89\textwidth] \RR^{G}{}_{C} \R^{D}{}_{A} \R^{F}{}_{E} F_{DG}{}^{E} F_{BF}{}^{C} - \R^{F}{}_{C} \left( \R^{E}{}_{D} F_{AF}{}^{D} + \R^{D}{}_{A} F_{FD}{}^{E}  \right)   F_{BE}{}^{C}\\
  - \RR^{E}{}_{C} F_{EA}{}^{D} F_{BD}{}^{C} = 0\,, \end{multlined} \\
&\RR^{C}{}_{F} \RR^{D}{}_{G} \R^{E}{}_{A} F_{DE}{}^{F} F_{BC}{}^{G} + 2 \RR^{C}{}_{D}\R^{E}{}_{A}  F_{FE}{}^{D} F_{BC}{}^{F} - c_G \R_{AB} = 0\,,\\
&\RR^{E}{}_{A} \R^{F}{}_{B} \R^{D}{}_{G} F_{DE}{}^{C} F_{CF}{}^{G} = \R^{E}{}_{B} \R^{C}{}_{D} F_{FA}{}^{D}F_{CE}{}^{F},\\
&\RR^{D}{}_{G} \R^{H}{}_{A} \R^{E}{}_{B}  F_{HD}{}^{C}F_{CE}{}^{G}  + \RR^{E}{}_{D} F_{BC}{}^{D}F_{AE}{}^{C} - c_G \RR_{AB}  - c_G \kappa_{AB} = 0\,, \\
&\RR^{C}{}_{G} \RR^{D}{}_{F} \R^{H}{}_{A} \R^{E}{}_{B}  F_{DE}{}^{G} F_{CH}{}^{F}  + 2\RR^{D}{}_{E} F_{BC}{}^{E}F_{AD}{}^{C} - c_G \RR_{AB} - 2 c_G \kappa_{AB} = 0\,, \\
& \RR^{G}{}_{A} \RR^{D}{}_{E} \R^{F}{}_{B}  F_{DG}{}^C F_{FC}{}^E - \RR^{D}{}_{F} \R^{E}{}_{C} F_{AD}{}^C F_{EB}{}^F +\R^{C}{}_{D} F_{CA}{}^E F_{BE}{}^D  = c_G \R_{AB}\,, \\
&\RR^{G}{}_{A} \RR^{D}{}_{E} \R^{F}{}_{C} F_{DG}{}^C F_{FB}{}^E = \RR^{D}{}_{F}\R^{E}{}_{C} F_{AE}{}^F F_{BD}{}^C,\\
&  \R^{C}{}_{A} \R^{D}{}_{B} F_{CD}{}^{B} = 0\,.
\end{align}
\endgroup

\section{Geometry in the non-orthonormal frame}\label{s:geometry}
Consider a general Riemannian target manifold $\mathcal{M}$ with local coordinates $x^{\mu}$ and endowed with a curved metric $G$. We work  in a frame formalism $\hat{e}_{A} = e_{A}{}^\mu \partial_\mu$ where the metric is constant but non-orthonormal:
\begin{equation}
G_{\mu\nu}(x) = e^{A}{}_\mu (x) G_{AB} e^{B}{}_\nu (x) \ . 
\end{equation}
 Requiring the spin-connection to be metric-compatible and torsion-free gives the following connection coefficients:
\begin{equation}\label{eq:conngeng}
\Gamma^{A}_{BC} = \frac{1}{2}G^{AE}\left(\Omega_{EB}{}^{D}G_{DC} + \Omega_{EC}{}^{D}G_{DB} \right) + \frac{1}{2}\Omega_{BC}{}^{A}\,,
\end{equation}
where $\Omega_{AB}{}^C$ are the anholonomy coefficients determined by,
\begin{equation}
\left[ \hat{e}_{A},\hat{e}_{B}\right] = \Omega_{AB}{}^{C}\hat{e}_C \,, \qquad  \Omega_{AB}{}^C = e_{A}{}^\mu e_{B}{}^\nu \left(\partial_\nu e^{C}{}_\mu -\partial_\mu e^{C}{}_\nu\right)\,.
\end{equation}

In our case, the target manifold is a Lie manifold $G$ endowed with a deformed geometry. Introducing  left-invariant one-forms $u=g^{-1}\mathrm{d}g = -i u^{A}_\nu T_A  d x^\nu  $ which satisfy,
\begin{equation}\label{eq:mcstreq}
\mathrm{d}u^{A}  = -\frac{1}{2} F_{BC}{}^A u^B \wedge u^C \,,
\end{equation}
we go to the frames $\hat{e}_A = u_A^\mu \partial_\mu$.
The deformed geometry in this frame is given by the constant non-orthonormal metric eq.~\eqref{eq:metricansatz} and by the torsion eq.~\eqref{eq:torsion},
 \be
 G_{AB} =  \alpha \kappa_{AB} +\gamma   \R^{2}_{AB} \ , \qquad H_{ABC}=  3 \beta F_{\left[AB\right.}{}^{D}\R_{\left.C\right]D} -k F_{ABC}  \ .
 \ee
The inverse metric is then (using $\R^{3} = -\R$):
\begin{equation}
G^{AB} = \frac{1}{\alpha}\kappa^{AB} + \frac{\gamma}{\alpha(\gamma - \alpha)}\RR^{AB}\,.
\end{equation}
For the spin-connection coefficients we find from eq.~\eqref{eq:mcstreq} that $\Omega_{AB}{}^B = F_{AB}{}^C$ and thus,
\begin{equation}\label{eq:spinconn}
\Gamma^{A}_{BC} = \frac{1}{2}G^{AE}\left(F_{EB}{}^{D}G_{DC} + F_{EC}{}^{D}G_{DB} \right) + \frac{1}{2}F_{BC}{}^{A}\,.
\end{equation}
Noting that the spin-connections are constant, the Riemann tensor can be calculated from,
\begin{equation}
R^{A}{}_{BCD} = \Gamma^{E}_{DB}\Gamma^{A}_{CE} - \Gamma^{E}_{CB}\Gamma^{A}_{DE}-\Omega_{CD}{}^{E}\Gamma^{A}_{EB}\,,
\end{equation}
and the Ricci tensor from,
\begin{equation}
R_{AB} = R^{C}{}_{ACB} = -\Gamma^{E}_{CA}\Gamma^{C}_{BE} -F_{CB}{}^{E}\Gamma^{C}_{EA}\,.
\end{equation}


With the $\beta$-functions in mind we end this appendix with a set of useful expressions which are found by plugging in the expressions of the metric eq.~\eqref{eq:metricansatz} and the torsion eq.~\eqref{eq:torsion} and by making use of the properties of the $\R$-matrix listed in appendix \ref{s:properties}:  
\begin{itemize}
\item The spin-connection: 
\begin{equation}\label{eq:conngenglie}
\Gamma^{A}_{BC} = \frac{1}{2}\frac{\gamma}{\alpha - \gamma}\left(F_{BD}{}^{A}\RR^{D}{}_{C} + F_{CD}{}^{A}\RR^{D}{}_{B} \right) + \frac{1}{2}F_{BC}{}^{A} \ . 
\end{equation}
\item The Ricci tensor: \begin{eqnarray}
R_{AB} &=& \frac{c_G}{4}\left( 1 + \left(\frac{\gamma}{\gamma-\alpha} \right)^{2}\right)\kappa_{AB} - \frac{c_G}{4}\left(1-\left(\frac{\alpha}{\gamma-\alpha}\right)^{2} \right) \R^{2}_{AB} \nonumber\\
&&
 +\frac{1}{2}\left(\frac{\gamma}{\alpha-\gamma}\right) F_{AD}{}^{C}F_{BC}{}^{E}\RR^{D}{}_{E}   \ . 
\end{eqnarray}
\item The Ricci curvature: \begin{equation}\label{eq:riccicurv}
R = R_{AB} G^{AB} = -\frac{c_G}{4} \left(\frac{1}{(\gamma-\alpha)}D +\frac{\gamma}{(\gamma - \alpha)^{2}}l\right) \ ,
\end{equation}
where $D$ is the dimension and $l$ is the rank of the Lie algebra $\mathfrak{g}$. Hence, we have $\mathrm{Tr} \RR = - (D-l)$.
\item Expressions from the torsion tensor: 
\begin{alignat}{2}
&H^{2}_{AB} &&= H_{ACD}H_{BEF}G^{CE}G^{DF}\nonumber\\
& &&=c_G \left( \frac{\alpha k^{2} - 2(\gamma k^{2} +\alpha\beta^{2})}{\alpha(\gamma - \alpha)^{2}} \right) \kappa_{AB} -c_G \frac{\beta^{2}}{(\gamma - \alpha)^{2}} \R^{2}_{AB}\nonumber\\& &&\qquad +\left( \frac{2(\gamma k^{2}+\alpha\beta^{2})}{\alpha(\gamma - \alpha)^{2}}\right)\RR^{D}{}_{E} F_{BC}{}^{E}F_{AD}{}^{C},\\
&H^{2} &&= -c_G \left( \frac{ k^{2}+\beta^{2}}{(\gamma-\alpha)^{3}}D -\frac{3(\gamma k^{2}+\alpha \beta^{2})}{\alpha(\gamma-\alpha)^{3}}l \right), \\
&\nabla_C H^{C}{}_{AB} &&= G^{DE} \left(\Gamma^{C}_{DA}H_{EBC} - \Gamma^{C}_{DB} H_{EAC} \right) \nonumber\\
& &&= c_G \frac{\beta}{\gamma - \alpha} \R_{AB} + \frac{2\beta}{\gamma-\alpha} \R^{D}{}_{E} F_{AD}{}^C F_{BC}{}^E .
\end{alignat}
\end{itemize}


\end{document}